\documentclass[sigconf]{acmart}

\usepackage[utf8]{inputenc}
\usepackage[T1]{fontenc}
\usepackage[english]{babel}
\usepackage{paralist}
\usepackage{xcolor}
\usepackage{hyperref}
\usepackage{amsmath}
\usepackage{graphicx}
\usepackage{amssymb}
\usepackage{amsthm}
\usepackage{hyperref}
\usepackage[capitalize,noabbrev]{cleveref}
\usepackage{tikz}
\usepackage{caption}
\usepackage{subcaption}
\usepackage{csquotes}
\usepackage{comment}

\usetikzlibrary{spline}
\usetikzlibrary{calc}

\newtheorem{observation}{Observation}
\crefname{section}{\S}{Sections}
\crefname{page}{page}{pages}
\crefname{paragraph}{\S}{Sections}
\Crefname{section}{Section}{Sections}
\crefformat{section}{\S#2#1#3}
\Crefformat{section}{Section~#2#1#3}

\renewcommand{\d}[1]{\,\mathrm{d}#1}

\def\BibTeX{{\rm B\kern-.05em{\sc i\kern-.025em b}\kern-.08emT\kern-.1667em\lower.7ex\hbox{E}\kern-.125emX}}
  
\setcopyright{none}
\acmConference[IFIP Performance]{IFIP Performance}{November 2--6, 
2020}{Milano, Italy}

\setcopyright{none}
\renewcommand\footnotetextcopyrightpermission[1]{} % Remove footer

\begin{document}

\title[Incentivizing Stable Path Selection]{%
%Toward Incentive-Compatible Traffic Engineering
%Enforcing Stable Path-Selection Strategies
Incentivizing Stable Path Selection\\in Future Internet Architectures
%\\in Path-Aware Network Architectures
}

\author{Simon Scherrer}
\affiliation{Department of Computer Science\\
ETH Zurich}

\author{Markus Legner}
\affiliation{Department of Computer Science\\
ETH Zurich}

\author{Adrian Perrig}
\affiliation{Department of Computer Science\\
ETH Zurich}

\author{Stefan Schmid}
\affiliation{Faculty of Computer Science\\
University of Vienna}

\begin{abstract}
    By delegating path control to end-hosts, future Internet architectures offer flexibility for path selection.
    However, there is a concern that the distributed
    routing decisions by end-hosts, in particular
    load-adaptive routing, can lead to oscillations
    if path selection is performed without coordination or
    accurate load information. Prior research has addressed this
    problem by devising path-selection policies that lead to
    stability.
    However, little is known about the viability of these policies
    in the Internet context, where selfish end-hosts can deviate
    from a prescribed policy if such a deviation is beneficial from
    their individual perspective. In order to achieve network
    stability in future Internet architectures, it is essential that
    end-hosts have an incentive to adopt a stability-oriented
    path-selection policy.

    In this work, we perform the first incentive analysis of the
    stability-inducing path-selection policies proposed
    in the literature.
    Building on a game-theoretic model of end-host path selection, we
    show that these policies are in fact incompatible with the
    self-interest of end-hosts, as these strategies make it worthwhile
    to pursue an oscillatory path-selection strategy.
    Therefore, stability in networks with selfish end-hosts must be
    enforced by incentive-compatible mechanisms. We present
    two such mechanisms and formally prove their incentive
    compatibility.
\end{abstract}

\maketitle

\vspace{-9pt}
\section{Introduction}
\label{sec:intro}

The past 20 years of research on next-generation Internet architectures have shown the benefits of path awareness and path control for end-hosts, and multiple path-aware network architectures have been proposed.
Many of these architectures, including RON~\cite{andersen2001resilient}, Platypus \cite{raghavan2004system}, MIRO \cite{xu2006miro}, Pathlets~\cite{godfrey2009pathlet}, Segment Routing~\cite{filsfils2015segment}, and SCION \cite{barrera2017scion}, allow end-hosts to select the inter-domain paths over which their data packets are forwarded.
One principal argument for such path control is that it enables load-adaptive routing, i.e., allows the end-hosts to avoid congested links, and should therefore lead to a relatively even traffic distribution.
However, load-adaptive routing creates new challenges, in particular the introduction of instabilities under certain conditions. 
%Already in ARPANET, the load-adaptive routing protocol produced unstable traffic patterns, before it was eventually abandoned in favor of a more static routing protocol~\cite{khanna1989revised}.
Instability due to load-adaptive routing typically appears in the form of \emph{oscillations}, i.e., periodic up- and downswings of link utilization, leading to a large variance of the traffic load in a short time span.
According to the IETF, a central obstacle to
deployment of path-aware network architectures are
`oscillations based on feedback loops, 
as hosts move from path to path'~\cite{irtf-panrg-what-not-to-do-07}.
Indeed, such oscillations can be shown to occur if path-selection decisions are taken on the basis of outdated load information~\cite{fischer2009adaptive, shaikh2001evaluating}, which is the case in any real system.

Such oscillations are undesirable for many reasons, both from the perspective of the end-hosts and the perspective of the network operator. 
If oscillation occurs when a link is near its capacity limit, there is a danger of queue build-up, jitter, and, as a result, unpredictable performance.
Moreover, oscillation temporarily leads to a heavily skewed load distribution over paths, causing higher overall queuing latency than with a more equal traffic distribution.
Due to the large variance of the load level over time, network operators have to perform substantial overprovisioning of link capacities, which is undesirable from a business perspective.
Moreover, oscillation of inter-domain traffic imposes additional overhead
for intra-domain traffic engineering (e.g., MPLS circuit setup), as oscillating inter-domain flows may constantly switch between inter-AS interfaces.
From the end-host perspective, oscillation causes packet loss and thus forces the congestion-control algorithms to recurring restarts, negatively affecting throughput.

To avoid these damaging effects, researchers have devised numerous schemes that aim to guarantee stability of load-adaptive routing. 
However, to the best of our knowledge, no scheme so far has aimed at providing stability in Internet architectures with end-host path control.
Many systems have been designed under the assumption of network-based path selection, i.e., hop-by-hop forwarding according to decisions taken by intermediate routers~\cite{fischer2006replex, gojmerac2003adaptive, kvalbein2009multipath, michael2014halo}.
These systems achieve convergence by appropriately adjusting how much traffic is forwarded to each next hop towards a destination and cannot be used if packets must be sent along paths selected by end-hosts. 
Other systems allow end-point path selection, but are targeted to an intra-domain context where the end-points (typically ingress and egress routers) are under the control of a network operator~\cite{elwalid2002mate, fischer2009adaptive, jonglez2017distributed, kandula2005walking, kelly2005stability, nelakuditi2002adaptive}. 
In an intra-domain context, network operators are able to prescribe arbitrary path-selection procedures that generate stability. 
Conversely, in an inter-domain context, the end-points are not under control of network operators and can thus not be forced to adopt a non-oscillatory path-selection strategy.
Instead, as end-hosts must be assumed to be selfish, they can only be expected to adopt path-selection strategies that optimize performance \emph{from their individual perspective}.

By performing a game-theoretic analysis, we show in this paper that the non-oscillatory path-selection strategies traditionally proposed in the literature on stable source routing~\cite{elwalid2002mate, fischer2009adaptive, jonglez2017distributed, kandula2005walking, kelly2005stability, nelakuditi2002adaptive} are incompatible with the self-interest of end-hosts.
Assuming that such non-oscillatory path-selection strategies are universally adopted, an end-host can increase its utility by deviating 
%from the non-oscillatory path-selection strategy 
in favor of a strategy that is oscillatory. 
Therefore, stability of load-adaptive routing in an inter-domain context cannot be achieved by relying only on end-point path selection.
Instead, network operators have to \emph{incentivize} end-hosts to adopt one of the well-known convergent path-selection strategies with \emph{stabilization mechanisms}.
These mechanisms have to be \emph{incentive-compatible}, i.e.,
the mechanisms must create an incentive structure such that it is
in an end-host's self-interest to adopt a non-oscillatory path-selection strategy.
In this work, we present two such stabilization mechanisms, FLOSS and CROSS, and formally prove their incentive compatibility. 
These mechanisms employ different
techniques to disincentivize
oscillatory switching between paths, namely limiting the
migration rate between paths (FLOSS) and imposing a cost on
switching between paths (CROSS).
To complement our mainly theoretical work,
we also discuss how our findings could be practically
applied.

\subsection{Contribution}
\label{sec:intro:contribution}

This paper revisits the theoretical study of the dynamic effects of end-point path selection, for the first time focusing the analysis on inter-domain networks where the end-points are selfish and
uncontrolled.
We present a game-theoretic model that allows us to investigate which path-selection strategies will be adopted by selfish end-hosts.
In particular, we introduce the notion of equilibria to path-selection strategies (PSS equilibria).
Moreover, we formally show that the non-oscillatory path-selection strategies proposed in the existing literature do not form such PSS equilibria.
Thus, we provide evidence towards the hypothesis that stability in load-adaptive routing over multiple domains cannot be achieved by exclusively relying on end-hosts' path-selection behavior.
To remedy this problem, we leverage insights from mechanism design to devise two incentive-compatible stabilization mechanisms enforced by network operators.
While these mechanisms build on existing 
insights from intra-domain traffic engineering, their methods of incentivization represent a novel approach to achieve stability in inter-domain networks with load-adaptive routing.
We formally prove the incentive compatibility of both mechanisms
and discuss their practical application.

% \subsection{Organization}
% \label{sec:intro:organization}

% The remainder of this paper is organized as follows.
% Building on a model introduced in \cref{sec:model}, we analyze 
% the non-oscillatory path-selection strategies from the literature 
% with respect to their 
% game-theoretic rationality in \cref{sec:strategies}.
% In \cref{sec:mechanisms}, we introduce the idea behind stabilization
% mechanisms and explain how such mechanisms can be enforced by network operators.
% Two traffic-engineering mechanisms by the name of FLOSS and CROSS are presented in \cref{sec:floss} and \cref{sec:cross}, respectively.
% We discuss possibilities for the practical application of
% our findings in \cref{sec:practical} and discuss related work in \cref{sec:related-work} before we conclude in \cref{sec:conclusion}.

\section{Oscillation Model}
\label{sec:model}

\subsection{Parallel-Path Systems}
\label{sec:model:system}

In order to study oscillation in network architectures with end-host path selection, we build on the well-established Wardrop model~\cite{wardrop1952road}, which is the standard model for studying the interactions of selfish agents in computer networks~\cite{ qiu2003selfish, roughgarden2003price, roughgarden2002bad}. 
In the Wardrop model, an infinite number of end-hosts, each controlling an infinitesimal traffic share, select one path $\pi$ among multiple paths $\Pi$ between two network nodes. 
Every path~$\pi$ has a load-dependent cost, where the path-cost function $c_{\pi}$ is typically interpreted as latency. 
The end-hosts' path-selection decisions form a congestion game, where the path-selection decisions of end-hosts both determine and follow the load $f_{\pi}$ on every 
path~$\pi$~\cite{christodoulou2005price, holzman1997strong, rosenthal1973class}.

In this work, we analyze congestion games with a temporal component, i.e., end-hosts take path-selection decisions over time based on currently available information. More precisely, an end-host performs an average of $r > 0$ re-evaluations per unit of time. The aggregate re-evaluation behavior is uniform over time, i.e., when dividing time into intervals of length $\epsilon \in (0,1]$, $r\epsilon$ re-evaluations are performed in any interval
% \footnote{An alternative interpretation of $r\epsilon$ is the average number of re-evaluations per end-host in an interval. We assume at most one re-evaluation per end-host in an interval and thus $r \leq 1/\epsilon$. However, as we will later choose $\epsilon$ to be infinitesimal, this assumption is no limitation of the model.}.

Whenever an end-host performs a re-evaluation, it chooses one path $\pi$ to its destination according to a freely chosen path-selection strategy $\sigma$. We thus formalize the environment of congestion games as \emph{parallel-path systems}:

\begin{definition}
    A \textbf{parallel-path system} $O:=(\Pi, r, p, T, A_0, v)$ is a tuple, where a total demand normalized to 1 is distributed over parallel paths $\pi \in \Pi$ among which end-hosts can select; $r > 0$ is the average number of re-evaluations per end-host and unit of time; $p \geq 1$ is the steepness of the path cost as a function of the load (i.e., $c_{\pi} = (f_{\pi})^p$); $T \geq 0$ is the average time that it takes for cost information to reach the agents; $\mathbf{A_0} \in \left[0,1\right]^{|\Pi|}$ is the initial load matrix, where the entry $\mathbf{A_0}_{\pi} = f_{\pi}(0)$; and $v$ is the strategy profile, defining for every available path-selection strategy $\sigma$ the share $v(\sigma)$ of end-hosts
    that permanently apply strategy $\sigma$.
\end{definition}

Every congestion game possesses at least one Wardrop equilibrium, consisting of a traffic distribution where no single agent can reduce its cost by selecting an alternative path~\cite{rosenthal1973class}. 
If the agents take path-selection decisions based on up-to-date cost information of paths ($T = 0$), convergence to Wardrop equilibria is guaranteed and persistent oscillations can thus not arise~\cite{fischer2004evolution, fischer2005evolutionary, sandholm2001potential}. 
However, in practice, the cost information possessed by agents is \emph{stale} ($T > 0$), i.e., the information describes an older state of the network.
If such stale information is present, undesirable oscillations can arise~\cite{fischer2009adaptive}. Therefore, parallel-path systems can be \emph{oscillation-prone}:

\begin{definition}
    A parallel-path system $O$ is \textbf{oscillation-prone} if and only if~$T > 0$.
\end{definition}

In this work, we study oscillation-prone systems with two paths $\alpha$ and $\beta$ (i.e., $|\Pi| = 2$), but our insights directly generalize to more paths.
Due to total demand normalization, it holds that $f_{\beta}(t) = 1 - f_{\alpha}(t)$ for all $t \geq 0$.
Thus, the unique Wardrop equilibrium in a two-path oscillation-prone system is given by $f_{\alpha}=f_{\beta}=1/2$.
Moreover, we assume w.l.o.g.\ that the initial imbalance $A_0$ exists with the higher load on path~$\alpha$: $f_{\alpha}(0) = A_0 = \mathbf{A_0}_{\alpha} > 1/2$.
For this system of two parallel paths, 
% we define 
% \begin{equation}
%     \tilde\pi = \begin{cases}
%         \beta,&\pi=\alpha,\\
%         \alpha,&\pi=\beta.
%     \end{cases}
% \end{equation}
$\tilde\pi$ denotes the respective other path, i.e., $\tilde\alpha=\beta$ and $\tilde\beta=\alpha$.

Having introduced the concept of oscillation-prone systems, we next define notions of oscillation and stability.
First, an oscillation-prone system experiences oscillation if the traffic distribution does not eventually become static:

\begin{definition}
    An oscillation-prone system $O$ experiences \textbf{oscillation} if there exists no limit $\Delta^{\ast}$ of the function $\Delta(t) = |f_{\alpha}(t)-f_{\beta}(t)|$ for $t \rightarrow \infty$.
\end{definition}

Conversely, we understand stability simply as the absence of oscillation, i.e., stability is given if a limit $\Delta^*$ exists.
However, to ensure optimal network utilization, the desirable state of the network is not only stability, but stability \emph{at equal load} as given by the Wardrop equilibrium:

\begin{definition}
    An oscillation-prone system $O$ is \textbf{stable at equal load} if $\Delta^* := \lim_{t\to\infty} \Delta(t) = 0$.
\end{definition}

\subsection{Path-Selection Strategies}
\label{sec:model:strategies}

In a congestion game, end-hosts select paths according to freely adopted path-selection strategies.
In order to enable a theoretical treatment, we follow Fischer
and V\"ocking \cite{fischer2009adaptive} in assuming that path-selection strategies are memory-less, i.e., not dependent on anything else than currently observable information.
Therefore, any path-selection strategy $\sigma$ can be fully characterized by two elements, $\sigma = (R, u)$, which we will describe in the following.

First, every strategy is characterized by the expected time~$R$ between re-evaluations of an end-host.
The expected re-evaluation period $R$ reflects the reallocation behavior of end-hosts that non-deterministically re-evaluate the costs of path options, decide for one option based on the perceived costs, and keep sending on the selected path until the next re-evaluation is due.
The expected re-evaluation period $R$ has to be in accordance with the parameter $r$ of the parallel-path system, which describes the average number of re-evaluations per end-host and unit of time.
Hence, $R = 1/r$.
%The expected time between two re-evaluations of a given end-host is $R=1/r$. 

Second, every strategy $\sigma$ is based on a path-selection function $u(\pi, t\mid \pi')$, which gives the probability for selecting path $\pi$ at time~$t$ if the currently used path is $\pi'$.
Given universal adoption of a strategy $\sigma$ and $r\epsilon$ re-evaluations per interval of length $\epsilon$, the number of end-hosts on path $\pi$ changes by the amount $\Delta_{\epsilon} f_{\pi}(t) = -r\epsilon\cdot u(\tilde\pi,t\mid \pi) \cdot f_{\pi}(t) + r\epsilon\cdot u(\pi,t\mid \tilde\pi)\cdot f_{\tilde\pi}(t)$ within an interval starting at time $t$, given a two-path system. If $\epsilon$ is chosen to be infinitesimal, we obtain the \emph{rate} of change:\begin{equation}
\begin{split}
    \frac{\partial f_{\pi}(t)}{\partial t} = \lim_{\epsilon\to 0} \frac{\Delta_{\epsilon} f_{\pi}(t)}{\epsilon} = &-r\cdot u(\tilde\pi,t\mid \pi) \cdot f_{\pi}(t)\\ &+ r\cdot u(\pi,t\mid \tilde\pi)\cdot f_{\tilde\pi}(t)
\end{split}
\end{equation} Throughout the rest of the paper, we describe oscillation dynamics by such differential equations.

An example of a path-selection strategy is the greedy path-selection strategy $\sigma_\mathrm{g}$, which selects the path perceived as cheaper: 
\begin{equation}
u_\mathrm{g}(\pi,t\mid \tilde\pi) = \begin{cases}1 & \text{if } c_{\pi}(t-T) < c_{\tilde\pi}(t-T)\\ 0 & \text{otherwise}\end{cases}
\end{equation} 
Conversely, the probability of staying on a path is $u_\mathrm{g}(\tilde\pi,t\mid \tilde\pi) = 1-u_\mathrm{g}(\pi,t\mid\tilde\pi)$. %\footnote{If both paths have equal costs, end-hosts stay with the current path.}
At time $t$, the number of end-hosts on a more expensive path~$\pi$ thus changes with rate $-r \cdot f_{\pi}(t)$.

Whether an oscillation-prone system in fact experiences oscillation entirely depends on the path-selection strategies adopted by end-hosts.
In the next section, we present the example of an oscillation-prone system that experiences oscillation for some path-selection strategy, but converges to stability for a different strategy.

\subsection{Example of Oscillation}
\label{sec:model:example}

For every $T>0$, oscillation occurs in a system in which all agents adopt a \emph{greedy} path-selection strategy $\sigma_\mathrm{g}$ presented in the previous section. 
The dynamics of a system with universal adoption of the greedy strategy are given by the partial differential equation:\footnote{An analogous equation holds for $f_\beta$.}
\begin{equation}
    \frac{\partial f_{\alpha}(t)}{\partial t} = 
        \begin{cases}
            -r \cdot f_{\alpha}(t) & \text{if } c_{\alpha}(t-T) > c_{\beta}(t-T)\\
            r \cdot f_{\beta}(t) & \text{if } c_{\alpha}(t-T) < c_{\beta}(t-T)\\
            0 & \text{otherwise}
        \end{cases}
 \end{equation} 
We henceforth refer to \emph{turning points} as all points in time $t^+$ where $c_{\alpha}(t^+-T) = c_{\beta}(t^+-T)$, as $f_{\alpha}(t)$ switches between increasing and decreasing at these moments, and write $t^+(t)$ for the most recent turning point $t^+ < t$. 

Solving the differential equation piece-wise yields the following recursive function:\footnote{In the two-path system, $f_\alpha\geq\frac12$ is equivalent to $c_\alpha\geq c_\beta$.}
\begin{equation}
        f_{\alpha}(t) = \begin{cases}
        e^{-r\cdot(t-t^+(t))}\cdot f_{\alpha}(t^+(t)) & \text{if } f_{\alpha}(t-T) \geq \frac12\\
        1 - e^{-r\cdot(t-t^+(t))}\cdot f_{\beta}(t^+(t)) & \text{otherwise}
    \end{cases}
\end{equation} 

\begin{figure}
    \centering
    \includegraphics[width=\linewidth]{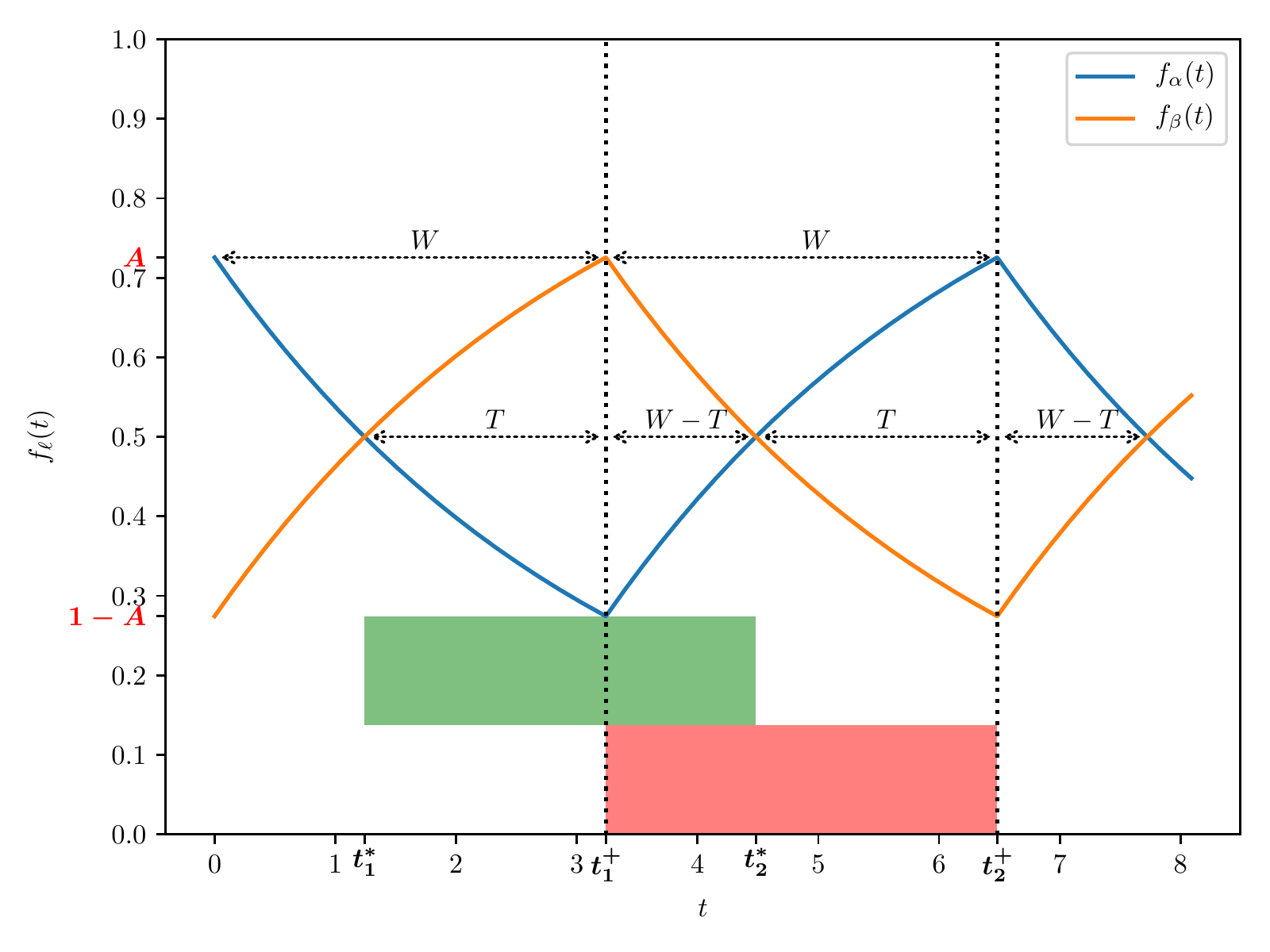}
    \vspace{-30pt}
    \caption{Oscillation structure for oscillation-prone system $O =\big(\{\alpha,\beta\},r=0.3,p\geq 1,T=2,A_0=A,v=\{\sigma_\mathrm{g}\mapsto1\}\big)$. $A$ and $W$ are calculated according to \cref{eq:AW}.}
    \label{fig:definition:oscillation-structure}
    \vspace{-10pt}
\end{figure}

Since~$T$ is constant, $f_{\alpha}(t)$ is periodic after the first turning point $t_1^+$ irrespective of the initial imbalance $A_0$. 
Therefore, the oscillation can be described by the non-recursive function:

\begin{equation}
f_{\alpha}(t) = \begin{cases}
e^{-r\cdot(t-t^+(t))}\cdot A & \text{if } \frac{t^+(t)}{W} \text{ is even,}\\
1 - e^{-r\cdot(t-t^+(t))}\cdot A & \text{otherwise},
\end{cases}
\end{equation} where 
\begin{equation}\label{eq:AW}
    W=\frac{\ln(2e^{rT}-1)}{r} , \quad A = 1 - \frac{1}{2e^{rT}},
\end{equation}
and $t^+(t) = t - (t \mod W)$ is a multiple of $W$.
\Cref{fig:definition:oscillation-structure} shows an example of $f_{\alpha}(t)$ for the oscillation-prone system $O = \big(\Pi=\{\alpha,\beta\},r=0.3,p\geq1,T=2,A_0=A,v=\{\sigma_\mathrm{g}\mapsto1\}\big)$, where $A_0$ has been chosen as $A$ in order to skip the irregular starting phase.
\Cref{fig:definition:oscillation-structure} also highlights the time interval during which path~$\alpha$ is the cheaper path (in green, between $t_1^{\ast}$ and $t_2^{\ast}$) and the time interval during which path $\alpha$ is perceived to be the cheaper path (in red, between $t_1^{+}$ and $t_2^{+}$).
Clearly, the discrepancy between reality and perception of path costs is the source of oscillation, as the discrepancy leads to increasing load on a path even when it is no longer the cheaper path (i.e., between $t_2^*$ and $t_2^+$).
Due to the periodicity of this phenomenon, there exists no limit $\Delta^{\ast}$ of load difference and the oscillation-prone system experiences oscillation.
An interesting observation is that both amplitude ($A$) and oscillation period ($2W$) increase with the staleness of the information ($T$); any $T>0$ leads to oscillations, only $T=0$ ensures stability.

In contrast, if the strategy profile~$v$
contains different path-selection
strategies, an oscillation-prone 
system may experience stability (cf.
example in Appendix~\ref{sec:model:example-stability}).

% \subsection{Example of Stability}
% \label{sec:model:example-stability}

% In contrast, the oscillation-prone system from the previous section is stable if a sufficient number of end-hosts anticipate the greedy strategy $\sigma_\mathrm{g}$ with an \emph{antagonist} strategy $\sigma_\mathrm{a}$.
% An end-host adopting the antagonist strategy always selects the path with the higher perceived cost, speculating that the seemingly cheaper path will soon be overloaded by greedy-strategy players: \begin{equation}u_\mathrm{a}(\pi,t\mid \tilde\pi) = \begin{cases}1 & \text{if } c_{\pi}(t-T) > c_{\tilde\pi}(t-T)\\ 0 & \text{otherwise}\end{cases}\end{equation}
% Conversely, $u_\mathrm{a}(\tilde\pi,t\mid \tilde\pi) = 1-u_\mathrm{a}(\pi,t\mid \tilde\pi)$.

% In an oscillation-prone system with strategy profile $v = \{\sigma_\mathrm{g} \mapsto q, \sigma_\mathrm{a} \mapsto 1-q\}$ and initial imbalance $A_0 > 1/2$, the initial dynamics of the system are \begin{equation}
% f_{\alpha}(t)=(A_0+q-1)e^{-rt} + (1-q).
% \end{equation} 

% For $q \leq 1/2$, we see that $f_{\alpha}(t) > f_{\beta}(t)$ for all $t \geq 0$, since $\lim_{t\rightarrow\infty} f_{\alpha}(t) = 1-q \geq 1/2$, $f_{\alpha}(0) = A_0 > 1/2$, and $f_{\alpha}(t)$ is monotonic. Using the definitions from \cref{sec:model:system}, the oscillation-prone system is \emph{stable} with $\Delta^* = 1-2q$ for all $q < 1/2$ and is \emph{stable at equal load} for $q = 1/2$.

\subsection{Equilibria on Path-Selection Strategies}
\label{sec:model:nash}

In general, Nash equilibria refer to strategy profiles that do not allow for beneficial selfish strategy changes by individual agents. In the context of path-selection strategies, a Nash equilibrium is thus given if every end-host cannot improve its utility by switching to an alternative path-selection strategy. More formally, a Nash equilibrium on path-selection strategies can be defined as follows: 

\begin{definition}
A strategy profile $v^*$ is a Nash equilibrium on path-selection strategies (\textbf{PSS equilibrium}) in an oscillation-prone system $O = (\Pi,r,p,T,A_0,v^*)$ if and only if all strategies $\sigma$ with $v^*(\sigma) > 0$ have cost $ C(\sigma\mid O) =~C^*$ and all strategies $\sigma'$ with $v^*(\sigma') = 0$ have cost $C(\sigma'\mid O) \geq~C^*$.
\end{definition}

It remains to formally define the cost $C(\sigma\mid O)$ of a strategy~$\sigma$ in an oscillation-prone system $O$ with global strategy profile~$v$. First, we note that a global strategy profile~$v$, together with an initial strategy-adoption distribution for each path, uniquely defines the flow dynamics~$f(t) = (f_{\alpha}(t),f_{\beta}(t))$ in oscillation-prone systems with two paths. As the flow share controlled by each agent is assumed to be negligible in the Wardrop model, the flow dynamics~$f(t)$ are not affected by the choice of $\sigma$ when varying $\sigma$ for a single agent. The basic costs of the two path options $\alpha$ and $\beta$ at any moment~$t$ are thus given by~$c_{\alpha}(t)$ and~$c_{\beta}(t)$, both uniquely defined by an oscillation-prone system $O = (\Pi,r,p,T,A_0,v)$.

Given expected re-evaluation periods of duration $R$, an end-host deciding for path $\pi$ at time $t$ incurs the \emph{usage cost}
\begin{equation}
c_\mathrm{u}(\pi,t) = \frac{1}{R}\int_t^{t+R}c_\pi(s)\d{s}. \label{eq:model:nash:usage-cost}
\end{equation}

At time $t$, the cost $c(\sigma,t)$ of applying a strategy $\sigma$ is
\begin{equation}
c(\sigma,t|\pi') = \sum_{\pi \in \Pi} u(\pi,t\mid \pi') \cdot c_\mathrm{u}(\pi,t),
\end{equation}
where $\pi'$ is the current path of the end-host before the decision at time $t$ and $u(\pi,t\mid \pi')$ is the probability that path $\pi$ is selected at time $t$ given the current path $\pi'$.

Furthermore, the strategy also determines the probability distribution $y(\pi'\mid t)$ that defines the probability of the current path being $\pi'$ at time~$t$. The expected cost for applying a strategy $\sigma$ at time $t$ is thus given as follows: \begin{equation} \begin{split} C(\sigma,t) &= \sum_{\pi' \in \Pi} y(\pi'\mid t) \cdot c(\sigma,t\mid \pi') \end{split}\end{equation}

The expected cost of applying a strategy $\sigma$ in general can be derived as the average time-dependent strategy cost during a certain \emph{relevant time span} $\big[t_0,t_1\big]$: 
\begin{equation}
C(\sigma\mid O) = \frac{1}{t_1-t_0} \int_{t_0}^{t_1}  C(\sigma,t) \d{t}
\end{equation} 
For systems that converge to stability at equal load, the relevant time span extends from $t_0=0$ until time $t_{\delta}$ when the system has converged according to some criterion $\delta > 0$, i.e., $\forall t > t_{\delta}.\ \Delta(t) < \delta$. The time after convergence does not have to be considered as all strategies have the same cost for a system with equal path costs.  For periodic oscillating systems, the relevant time span is defined as every interval that contains the periodically repeated sub-function. For an example of a PSS equilibrium analysis, see \cref{sec:model:equilibrium:test}.

\section{Limits of Stable Strategies}
\label{sec:strategies}

In this section, we investigate whether the stability-inducing path-selection strategies proposed in the literature form PSS equilibria. The question is whether an end-host can minimize its cost with a stability-oriented strategy if that strategy is universally adopted. 

We perform this investigation by means of two case studies.
In \cref{sec:strategies:fischer}, we analyze the convergent rerouting policies designed by Fischer and V\"ocking \cite{fischer2009adaptive} and show that such rerouting policies are not compatible with the selfishness of end-hosts. 
In \cref{sec:strategies:mate}, we analyze the MATE algorithm \cite{elwalid2002mate} and show its equivalence to the rerouting policies discussed in \cref{sec:strategies:fischer}. 

\subsection{Rerouting Policies by Fischer \& V\"ocking}
\label{sec:strategies:fischer}

A typical example of a convergent path-selection strategy has been proposed by Fischer and V\"ocking \cite{fischer2009adaptive}. 
The proposed path-selection strategy, which we henceforth refer to as the convergent strategy $\sigma_\mathrm{c}$, works as follows: If an end-host discovers a path with lower cost according to stale information, the end-host switches to that path with a probability that is a linear function of the perceived latency difference. 
More formally, the probability $u(\pi,t\,|\,\tilde\pi)$ to switch from path $\pi_t$ to path $\pi \neq \tilde\pi$ at time $t$ is:\begin{equation}
    u(\pi,t\,|\,\tilde\pi) = \begin{cases}\mu \cdot \frac{c_{\tilde\pi}(t-T) - c_{\pi}(t-T)}{\Delta_{\max}} & \text{if }c_{\pi}(t-T) < c_{\tilde\pi}(t-T),\\ 0 & \text{otherwise},  \end{cases} \label{eq:strategies:switching-probability-fischer}
\end{equation} 

Here, $\mu$ is a parameter in $[0,1]$ and the latency difference is normalized by~$\Delta_{\mathrm{max}}$, which is 1 in parallel-path systems
as defined in \cref{sec:model:system}. 
The dynamics of a two-path oscillation-prone system where strategy $\sigma_\mathrm{c}$ is universally adopted can thus be described by the  delay-differential equation (DDE) 
\begin{equation}
\frac{\partial f_{\alpha}}{\partial t} = \begin{cases} r\cdot\mu\cdot\Delta c(t-T)\cdot f_{\alpha}(t) & \text{if } 
\Delta c(t-T) \leq 0,\\
r\cdot\mu\cdot\Delta c(t-T)\cdot f_{\beta}(t) & \text{otherwise},
\end{cases}
\label{eq:strategies:dde-fischer}
\end{equation} where $\Delta c(t-T) = c_{\beta}(t-T) - c_{\alpha}(t-T)$.
This DDE describes a damped oscillator with delayed feedback and does not have an explicit solution~\cite{campbell1995complex}. 
However, we can numerically compute a solution using the method of steps \cite{erneux2009applied}.

\begin{figure}
    \centering
    \includegraphics[width=\linewidth]{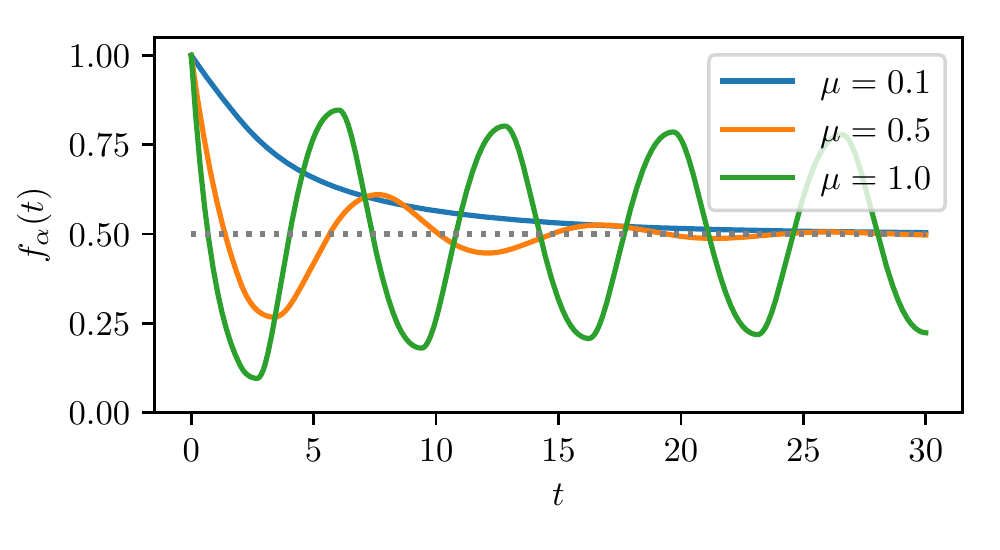}
    \vspace{-30pt}
    \caption{Dynamics produced by universal adoption of strategy $\sigma_\mathrm{c}$ with different $\mu$ in oscillation-prone system $O = (\{\alpha,\beta\},r=1,p=1,T=2,A_0=1,v=\{\sigma_\mathrm{c}\mapsto 1\})$.}
    \label{fig:strategies:fischer-strategies}
    \vspace{-13pt}
\end{figure}

As \cref{fig:strategies:fischer-strategies} shows, the choice of the parameter $\mu$ is critical for the strategy to actually lead to convergence. For high values of $\mu$, such as 1, the strategy fails to produce convergence and yields undamped periodic oscillations. For low values of $\mu$, such as 0.1, the system monotonically approaches the equilibrium without overshooting, i.e., it is \emph{overdamped} (or, if nearly avoiding overshooting, \emph{critically damped}). For values in-between, such as 0.5, the system eventually converges to stability at equal load, but only after overshooting, i.e., it is \emph{underdamped}. However, for both the overdamped and the underdamped convergent strategies, we can make the following observation:

\begin{observation}
    Universal adoption of the convergent path-selection strategy $\sigma_{\mathrm{c}}$ does not represent a PSS equilibrium, neither in its underdamped nor in its overdamped variant.
    \label{obs:strategies:fischer}
\end{observation}

In the case of the overdamped strategy (e.g., $\sigma_\mathrm{c}$ with $\mu = 0.1$), the link loads monotonically approach each other and thus the greedy strategy allows an end-host to make use of a cheaper path sooner, making it the best-response strategy given universal adoption of~$\sigma_\mathrm{c}$. In the case of the underdamped convergent strategy (e.g., $\sigma_\mathrm{c}$ with $\mu = 0.5$), the fact that the strategy is not a PSS equilibrium in general is not obvious. However, we can show that there exist alternative strategies to the underdamped rerouting policy that reduce a deviant agent's cost, see \cref{sec:appendix:obs-fischer}. 

\subsection{MATE Algorithm}
\label{sec:strategies:mate}

The MATE algorithm \cite{elwalid2002mate} was designed for the 
intra-domain context, where an ingress router has to distribute
its demand $d$ between multiple label-switched paths
to a given egress router. As these ingress routers are under control of the domain operator, the MATE algorithm pursues convergence to 
the socially optimal traffic distribution, which minimizes latency 
from a global perspective, but is generally unstable given selfish 
end-hosts. 
In the context of inter-domain networks, the MATE algorithm is 
instantiated such that it converges to a Wardrop equilibrium, 
a type of equilibrium that is stable under the assumption of selfish agents.

We analyze whether applying the MATE algorithm is rational from an end-host's perspective. An end-host in an oscillation-prone two-path system would execute 
the MATE algorithm as follows. 
In every re-evaluation, the end-host selfishly optimizes its 
traffic allocation $\big(F_{\alpha},F_{\beta}\big)^{\top}$, 
where $F_{\alpha} = d - F_{\beta}$. In order to conform to the 
Wardrop model, the demand $d$ is negligible from a global 
perspective. A MATE optimization step is defined as follows: \begin{equation}
\begin{pmatrix}F_{\alpha}'\\F_{\beta}'\end{pmatrix} = \left[\begin{pmatrix}F_{\alpha}\\F_{\beta}\end{pmatrix} - \gamma \cdot \begin{pmatrix}c_{\alpha}(t-T)\\c_{\beta}(t-T)\end{pmatrix}\right]^+
\end{equation} 
In order to reach convergence despite stale information, the coefficient $\gamma$ has to conform to a certain upper bound \cite{elwalid2002mate}. Moreover,~$\big[\mathbf{F}\big]^+$ represents a projection of allocation vector $\mathbf{F}$ to the feasible allocation set defined by $F_{\alpha} + F_{\beta} = d$ with $F_{\alpha}, F_{\beta} \geq 0$.

As we show in Appendix \ref{sec:appendix:obs-mate}, the dynamics of an oscillation-prone system with universal adoption of the MATE algorithm are described by the following differential equation: \begin{equation}\frac{\partial f_{\alpha}}{\partial t} = \begin{cases} r\cdot \frac{\gamma}{2} \cdot \Delta(t-T) \cdot f_{\alpha}(t) & \text{if } \Delta(t-T) \leq 0 \\ r\cdot \frac{\gamma}{2} \cdot \Delta(t-T) \cdot f_{\beta}(t) & \text{otherwise} \end{cases} \end{equation}  This equation is clearly equivalent to \cref{eq:strategies:dde-fischer} for a choice of $\mu = \gamma/2$. An oscillation-prone system with universal adoption of $\sigma_\mathrm{c}$ and a system with universal adoption of the MATE algorithm thus exhibit the same flow dynamics, which allow for beneficial deviation:

\begin{observation}
    The path-selection strategy as prescribed by the MATE algorithm is equivalent to the path-selection strategy $\sigma_{\mathrm{c}}$. Thus, universal adoption of the MATE algorithm neither constitutes a PSS equilibrium.
    \label{obs:strategies:mate}
\end{observation}

\subsection{Conclusion}
\label{sec:strategies:conclusion}

In summary, the kind of convergent path-selection strategies proposed in the literature cannot be assumed to be adopted by selfish end-hosts, as deviating from these strategies (e.g., by switching faster than prescribed by the strategy) is beneficial to an end-host.

Stability in a path-aware network architecture with selfish 
end-hosts can thus not be guaranteed by non-oscillatory path-selection strategies that prescribe a maximum rate of change to be respected by end-hosts. 
Instead, the network could employ \emph{mechanisms} that 
incentivize end-hosts to follow non-oscillatory path-selection 
strategies. This finding reflects a similar
result~\cite{akella2002selfish, godfrey2010incentive} in the 
context of congestion control, namely that socially desirable 
behavior of end-hosts can only be enforced with network support. 
\section{Stabilization Mechanisms}
\label{sec:mechanisms}

As argued in the previous section, rational end-hosts in networks 
with unrestricted path choice are unlikely to adopt convergent 
path-selection strategies. Therefore, there is a need for mechanisms 
that allow network operators to incentivize the adoption of 
path-selection strategies that induce stability at equal load, i.e.,
\emph{incentive-compatible} stabilization mechanisms. 
First, we integrate the concept of traffic-steering mechanisms into
our game-theoretic model (\cref{sec:mechanisms:steering}). 
Second, we specify in
\cref{sec:mechanisms:incentive-compatibility} the conditions
under which these mechanisms are incentive-compatible.

\subsection{Traffic-Steering Mechanisms}
\label{sec:mechanisms:steering}

In order to affect the path-selection decisions of end-hosts in an
oscillation-prone system~$O$, a traffic-steering mechanism 
$\mathcal{M}$ needs to alter the strategy cost $C(\sigma|O)$ for at 
least one path-selection strategy $\sigma$. A mechanism $\mathcal{M}$
thus defines a function $c_{\mathcal{M}}(\pi,t)$ that quantifies the 
mechanism-imposed cost for using path $\pi$ at time $t$. This cost is
imposed onto the user of a path $\pi$ in addition to the 
load-dependent path cost.

If a mechanism $\mathcal{M}$ is active, the usage 
cost~$c_{\mathrm{u}}^{\mathcal{M}}$ extends the standard usage 
cost~$c_{\mathrm{u}}$ from Equation (\ref{eq:model:nash:usage-cost}) 
as follows: \begin{equation}c_{\mathrm{u}}^{\mathcal{M}}(\pi,t) = 
c_{\mathrm{u}}(\pi,t) + c_{\mathcal{M}}(\pi,t)\end{equation}
The cost formulas 
$c^{\mathcal{M}}(\pi,t|\tilde\pi)$, $C^{\mathcal{M}}(\sigma,t)$, and 
$C^{\mathcal{M}}(\sigma|O)$ can be constructed from $c_{\mathrm{u}}^{\mathcal{M}}(\pi,t)$, analogously to
\cref{sec:model:nash}.

\subsection{Incentive Compatibility}
\label{sec:mechanisms:incentive-compatibility}

In general, incentive-compatible mechanisms are mechanisms that 
incentivize a certain form of desirable behavior. In our context, we 
consider traffic-steering mechanisms to be incentive-compatible if 
these mechanisms incentivize the desirable behavior of adopting a 
non-oscillatory path-selection strategy. In other words, an 
incentive-compatible mechanism creates a PSS equilibrium,
i.e., a situation where every end-host minimizes its cost
by adopting a non-oscillatory path-selection strategy,
given that all other end-hosts do so:

\begin{definition}
A \textbf{traffic-steering mechanism} $\mathcal{M}$ is an
incentive-compatible stabilization mechanism
for an oscillation-prone system~$O$ if
there is a strategy profile~$v^*$ such that
\begin{enumerate}[(i)]
    \item $v^*$ leads to stability at equal load and
    \item $v^*$ represents a PSS~equilibrium with respect to the cost function~$C^{\mathcal{M}}(\sigma|O)$.
\end{enumerate}
\label{def:inc-comp}
\end{definition}

In the following two sections, we present two instances of
stabilization mechanisms, namely FLOSS and CROSS, 
and prove their incentive compatibility.
The two mechanisms differ in the methods for achieving
stability: Whereas FLOSS reduces the imbalance between two
paths by regulating the migration rate between the paths,
CROSS achieves stability by repetitive reshuffling of
flows between paths and increasing the cost of path migration.
\section{The FLOSS Mechanism}
\label{sec:floss}

In this section, we present the FLOSS mechanism (\emph{Flow-Loyalty Oscillation-Suppression System}). %
%Finally, in \cref{sec:floss:practical}, we specify how the mechanism could be practically implemented such that the overhead for network operators is manageable.

\subsection{Overview}
\label{sec:floss:overview}

As shown in \cref{sec:strategies}, convergent path-selection 
strategies are characterized by careful path-switching behavior: An 
end-host only switches to a seemingly cheaper path with a modest 
probability that depends on the measured latency difference, translating into a relatively low migration rate between paths. It is well known that system
stability can be achieved by
by limiting the rate of change (also
known as the \emph{system gain}~\cite{kelly2005stability}).
However, the challenge is to
develop methods that achieve this
change-rate limitation in the
face of selfish, uncontrolled end-hosts.
Such a method is given by FLOSS.

As selfish end-hosts do not voluntarily conform to a modest 
path-migration rate, the path-migration rate has to be 
\emph{regulated} by network operators. The FLOSS mechanism performs 
such regulation by rewarding end-hosts that are \emph{loyal} to a 
certain path and by restricting arbitrary path migration by 
oscillating end-hosts.

\begin{figure}
    \centering
    \includegraphics[width=\linewidth]{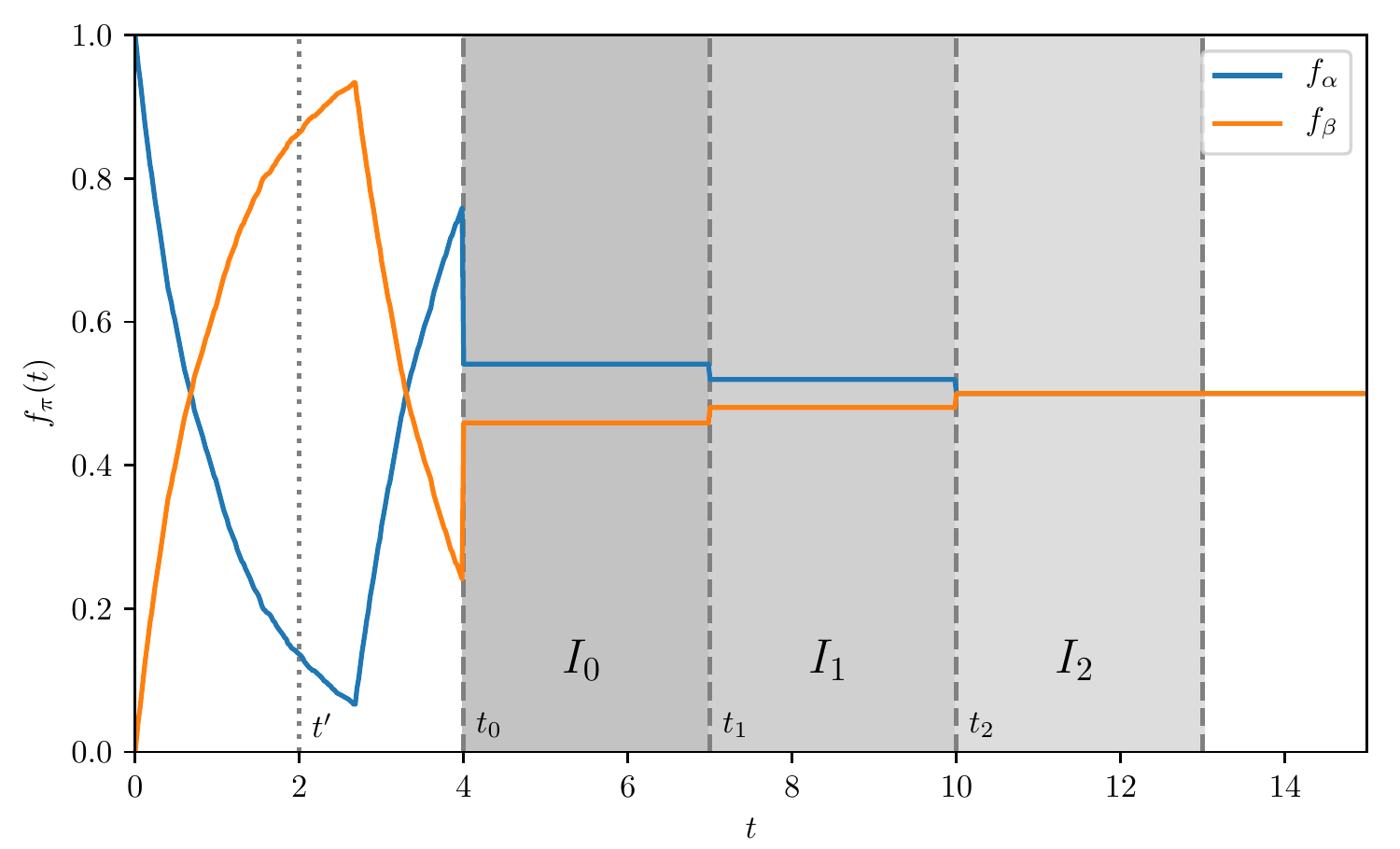}
    \vspace{-25pt}
    \caption{Simulation of FLOSS enforcement in an oscillation-prone system $O = (\Pi=\{\alpha,\beta\},r=1,p=1,T=2,A_0=1,v=\{\sigma_{\mathrm{F}}\mapsto1\})$.}
    \label{fig:floss:simulation}
    \vspace{-15pt}
\end{figure}

In order to regulate path migration, the FLOSS mechanism makes
use of \emph{registrations} and proceeds in \emph{intervals}. Figure \ref{fig:floss:simulation}, which shows 
a simulation of the FLOSS mechanism in a
two-path system, illustrates the FLOSS approach.
Initially, the FLOSS mechanism announces at time $t'$ that
all end-hosts are required to obtain a registration for 
one path $\pi$ of their choice. 
This registration allows an end-host to use path $\pi$
during a future time interval $I_0=[t_0, t_1)$ with 
$t' < t_0 < t_0 + T < t_1$.
End-hosts that use path~$\pi$ without a registration
are punished in the interval (e.g., by dropping packets).

This call for registration produces a distribution of flows
over the two paths, which is stable during the interval as
no end-host can switch to the path which it is not registered for.
However, this load distribution is unlikely to be 
perfectly equal.
The FLOSS mechanism iteratively reduces this imbalance:
In every following time interval, a small set of flows are
allowed to migrate from the more
expensive path to the cheaper path. This allowance is enforced
by selectively granting registrations: Whereas end-hosts
with a pre-existing registration for a path (loyal end-hosts)
always obtain a registration for that path, end-hosts
without a pre-existing registration are not always allowed
to register. Once the imbalance is sufficiently small,
the end-hosts do not have an incentive anymore to switch
paths, at which point the enforcement of the mechanism
can be suspended (e.g., at the end of interval $I_2$ in
Figure~\ref{fig:floss:simulation}).

\begin{theorem}
The FLOSS mechanism is an incentive-compatible stabilization
mechanism.
\label{thm:floss}
\end{theorem}

As defined in \cref{sec:mechanisms:incentive-compatibility},
incentive compatibility implies the existence of a strategy
profile that leads to stability at equal load \emph{and}
is a PSS equilibrium during mechanism enforcement.
For FLOSS, such a strategy profile is given by universal
adoption of the FLOSS-compliant path-selection strategy $\sigma_{\mathrm{F}}$. The strategy $\sigma_{\mathrm{F}}$
prescribes to use the path with the lowest expected cost 
which the end-host is entitled to use. Our
incentive-compatibility proof thus builds on the following
two concrete lemmas, which are proved in \cref{sec:floss:stability}
and \cref{sec:floss:equilibrium}, respectively:

\begin{lemma}
Universal adoption of the FLOSS path-selection 
strategy~$\sigma_{\mathrm{F}}$ leads to stability at
equal load.
\label{lem:floss:stability}
\end{lemma}

\begin{lemma}
Universal adoption of the FLOSS path-selection 
strategy~$\sigma_{\mathrm{F}}$ represents a PSS equilibrium during
enforcement of the FLOSS mechanism.
\label{lem:floss:equilibrium}
\end{lemma}

\subsection{Stability Analysis}
\label{sec:floss:stability}

In order to prove \cref{lem:floss:stability}, 
we assume universal adoption of path-selection strategy
$\sigma_{\mathrm{F}}$, i.e., an end-host always uses the
path with the lower expected cost provided that the end-host
is entitled to use that path.

When registering before the initial interval,
all end-hosts simultaneously decide for one path to use during
the upcoming interval $[t_0, t_1)$.
Confronted with such a choice, each end-host aspires to
commit to the path $\pi$ that will be selected by 
fewer other end-hosts, i.e.,
the path $\pi$ with $f_{\pi}(t_0) < f_{\tilde\pi}(t_0)$.
In absence of inherent differences between the two choices,
the only Nash equilibrium of such a speculative game is given if
every end-host commits to each path~$\pi$ with probability~$1/2$.

In expectation, the load on both paths $\alpha$ and $\beta$
is thus $\mathbb{E}[f_{\alpha}(t_0)] = \mathbb{E}[f_{\beta}(t_0)] 
= 1/2$. Since no migration occurs during the interval $[t_0,t_1)$, 
the load distribution is expected to remain equal during the interval, i.e.,
$\mathbb{E}[f_{\alpha}(t)] = \mathbb{E}[f_{\beta}(t)] = 1/2$ $\forall t \in [t_0,t_1)$.
When mechanism enforcement ends at time $t_1$, the end-hosts
are again free to arbitrarily select paths.
However, since $t_0 + T < t_1$, 
any end-host performing a re-evaluation
after $t_1$ perceives the Wardrop equilibrium 
$c_{\alpha}(t-T) = c_{\beta}(t-T)$ and will thus not switch paths.
Therefore, the system is stable at equal load even when 
the mechanism is not enforced anymore.

In reality, however, variance makes it likely that
the load on paths~$\alpha$ and $\beta$ is not perfectly
equalized at $t_0$. In that case, the FLOSS mechanism attempts to
eliminate the remaining load difference
$\Delta(t_0) = |f_{\alpha}(t_0) - f_{\beta}(t_0)| > 0$ as follows.
Starting from $t'' = t_0 + T$, the
end-hosts can again register on paths for an upcoming interval
$[t_1, t_2)$. At $t''$, all end-hosts correctly perceive
the cost difference between a cheaper path~$\pi$ and a more 
expensive path~$\tilde\pi$, as for every path~$\hat{\pi}$,
$c_{\hat{\pi}}(t''-T) = c_{\hat{\pi}}(t_0) = c_{\hat{\pi}}(t'')$
due to the constant load in $[t_0,t'')$. 
The core idea of the FLOSS mechanism is 
to determine and enforce a \emph{migration allowance}~$\rho_{\pi}(t_1)$,
which is an upper bound on the amount of end-hosts that
are allowed migrate from path~$\tilde\pi$ to path~$\pi$ at time~$t_1$.

Importantly, $\rho_{\pi}(t_1)$ is chosen such that
\begin{equation}f_{\pi}(t_0) + \rho_{\pi}(t_1)\cdot f_{\tilde\pi}(t_0) \leq 
(1-\rho_{\pi}(t_1))\cdot f_{\tilde\pi}(t_0),\end{equation} 
which implies $c_{\pi}(t_1) \leq c_{\tilde\pi}(t_1)$ 
(i.e., the cheaper path $\pi$ will remain the cheaper path in the
next interval even if a share $\rho_{\pi}(t_1)$ of end-hosts on the 
more expensive path~$\tilde\pi$ migrate to path~$\pi$). This choice 
of~$\rho_{\pi}(t_1)$ ensures the correct incentives for the 
end-hosts. Given such an assurance, end-hosts registered on the 
cheaper path $\pi$ during $[t_0, t_1)$ minimize their cost by 
remaining on path~$\pi$.
Since these end-hosts are considered \emph{loyal} to path $\pi$,
their registration at path~$\pi$
will be renewed for the upcoming interval $[t_1,t_2)$.
Conversely, all end-hosts registered on the 
more expensive path~$\tilde\pi$
would minimize their cost by migrating to the cheaper path~$\pi$.
However, the FLOSS mechanism restricts this migration by 
only granting a registration for~$\pi$ to a share~$\rho_{\pi}(t_1)$ 
of end-hosts on $\tilde\pi$. The non-migrating end-hosts
on path~$\tilde\pi$ are considered loyal on path~$\tilde\pi$ and 
are thus allowed to renew their registration at~$\tilde\pi$. 

Therefore, exactly $\rho_{\pi}(t_1)\cdot f_{\tilde\pi}(t_0)$ 
migrate from path~$\tilde\pi$ to path~$\pi$ at time~$t_1$, 
which reduces the difference in load and cost between the 
paths~$\pi$ and~$\tilde\pi$.
By repetitive mechanism application
with appropriately chosen migration allowances, the FLOSS mechanism
can arbitrarily minimize the cost differential between the 
paths~$\pi$ and~$\tilde\pi$. 
When the cost difference becomes so small 
that end-hosts perceive a Wardrop equilibrium, 
the mechanism has achieved stability at equal load that continues
to hold even without mechanism enforcement.

\subsection{PSS Equilibrium Analysis}
\label{sec:floss:equilibrium}

We now prove \cref{lem:floss:equilibrium}, i.e., we show that 
path-selection strategy $\sigma_{\mathrm{F}}$
is the optimal strategy for an end-host given that all other
end-hosts have adopted $\sigma_{\mathrm{F}}$.
Concretely, we show that the FLOSS mechanism induces a 
PSS equilibrium $v^* = \{\sigma_{\mathrm{F}}\mapsto1\}$, where 
$\sigma_{\mathrm{F}}$ is the universally adopted path-selection 
strategy with the following path-selection 
function:

\begin{equation}u_{\mathrm{F}}(\pi,t|\tilde\pi) = 
\begin{cases}
1/2 & \text{if } t = t_0,\\
1 & \text{if } \parbox[t]{.5\linewidth}{$t > t_0$ and $E_{e}(\pi,t)$ \\and $c_{\pi}(t-T) < c_{\tilde\pi}(t-T)$,}\\ 0 & \text{otherwise}\end{cases} \label{eq:floss:proof:u}\end{equation} 
where $E_e(\pi,t)$ is true if and only if end-host $e$ is entitled to use path~$\pi$ at time $t$. We assume that an end-host
always knows whether it is entitled to use
a path.
For the initial interval, every path is selected with equal
probability~1/2. For all subsequent intervals, a path $\pi$ is 
selected if the path is perceived to be cheaper
than the current path~$\tilde\pi$ and end-host $e$ is entitled to use path~$\pi$. For remaining on a path~$\tilde\pi$, it holds that $u_{\mathrm{F}}(\tilde\pi,t|\tilde\pi) = 1 - u_{\mathrm{F}}(\pi,t|\tilde\pi)$.

The FLOSS mechanism makes strategy $\sigma_{\mathrm{F}}$ the 
equilibrium strategy by
imposing the additional cost $c_{\mathcal{M}}(\pi,t)$ for using
path $\pi$ at time $t$.
End-host $e$ incurs a cost $c_{\mathrm{a}}$ for attempting
to register and a penalty cost $c_{\mathrm{p}}$ for using a path
without a registration. We assume $c_{\mathrm{p}} = \infty$, i.e.,
the penalty cost makes a path unusable.
Let~$A_e(\pi,t)$ be true 
if and only if end-host~$e$ applies 
to register for using path $\pi$ at time~$t$ 
and let~$R_e(\pi,t)$ be true if and only if end-host~$e$ obtained a registration for using path $\pi$ at time~$t$, i.e., 
$R_e(\pi,t) = A_e(\pi,t) \land E_e(\pi,t)$.
Using these predicates, the cost imposed by the FLOSS mechanism can be
expressed as \begin{equation}c_{\mathcal{M}}(\pi,t|A_e,R_e) = [A_e(\pi,t)]\cdot c_{\mathrm{a}} + [\neg R_e(\pi,t)]\cdot c_{\mathrm{p}},\end{equation} where $[P] = 1$ if the predicate $P$ is true and 0 otherwise.

A selfish end-host~$e$ chooses its actions such that its cost
from the mechanism is minimized. Therefore, an end-host $e$ requests
a registration if and only if the end-host is entitled to the 
registration, as there is no benefit of a registration request
that will be refused. Thus the relevant mechanism-imposed
cost for end-host $e$ is \begin{equation}c_{\mathcal{M}}(\pi,t) = \min_{A_e,R_e} 
c_{\mathcal{M}}(\pi,t|A_e,R_e) = \begin{cases}c_{\mathrm{a}} & 
\text{if } E_e(\pi,t),\\ c_{\mathrm{p}} & 
\text{otherwise.}\end{cases}\end{equation}

Concerning the initial interval with start~$t_0$,
both paths~$\alpha$ and~$\beta$ have expected cost $c_{\pi}(t_0) = 1/2^p$
if all other end-hosts choose each path with probability $u_{\mathrm{F}}(\pi,t|\tilde\pi) = 1/2$. 
As both paths have the same cost and both paths require
a registration, the usage cost of both paths is $c_{\mathrm{u}}^{\mathcal{M}}(\pi,t_0) = 1/2^p + c_{\mathrm{a}}$. 
Independent of the current path $\tilde\pi$, the cost of applying
strategy $\sigma_{\mathrm{F}}$ at time $t_0$ is thus
$c^{\mathcal{M}}(\sigma_{\mathrm{F}},t_0|\tilde\pi) = 1/2^p + c_{\mathrm{a}}$ for any
choice of $u(\pi,t_0|\tilde\pi)$. Therefore, end-host~$e$ cannot reduce
its cost by choosing another path-selection probability than $u_{\mathrm{F}}(\pi,t_0|\tilde\pi) = 1/2$, which makes $\sigma_{\mathrm{F}}$
an equilibrium strategy for the initial interval.

Concerning subsequent intervals with start $t_i > t_0$,
we have to distinguish two cases for the current path~$\pi'$ of
end-host $e$, namely whether end-host $e$ is on the
cheaper path~$\pi$ or on the more expensive 
path~$\tilde\pi$.\footnote{Thanks 
to the load being constant in subsequent intervals, 
the cost $c_{\hat{\pi}}(t)$ of a path $\hat{\pi}$ at registration 
time $t$ is equal to the
known stale cost $c_{\hat{\pi}}(t-T)$. Therefore, any end-host
can correctly identify the cheaper and the more expensive path.}
\begin{enumerate}
\item If end-host $e$ is on the cheaper path $\pi$, the cost of remaining on~$\pi$ is 
$c_{\mathrm{u}}^{\mathcal{M}}(\pi,t_i) = c_{\pi}(t_i) + c_{\mathrm{a}}$, 
whereas the cost of switching to~$\tilde\pi$ is 
$c_{\mathrm{u}}^{\mathcal{M}}(\tilde\pi,t_i) = c_{\tilde\pi}(t_i) + c_{\mathrm{a}}$
if $E_e(\tilde\pi,t_i)$ and $c_{\tilde\pi}(t_i) + c_{\mathrm{p}}$ otherwise.
As always $c_{\mathrm{u}}^{\mathcal{M}}(\pi,t_i) < c_{\mathrm{u}}^{\mathcal{M}}(\tilde\pi,t_i)$, the current path 
$\pi$ must be selected with probability $u(\pi,t|\pi) = 1$
to minimize the end-host's cost. 
\item If end-host $e$ is on the
more expensive path $\tilde\pi$, the cost of remaining on $\tilde\pi$ is $c_{\mathrm{u}}^{\mathcal{M}}(\tilde\pi,t_i) = c_{\tilde\pi}(t_i) + c_{\mathrm{a}}$, whereas the cost of switching to~$\pi$ is $c_{\mathrm{u}}^{\mathcal{M}}(\pi,t_i) = c_{\pi}(t_i) + c_{\mathrm{a}}$
if $E_e(\pi,t_i)$ and $c_{\pi}(t_i) + c_{\mathrm{p}}$ otherwise.
Thus, $c_{\mathrm{u}}^{\mathcal{M}}(\pi,t_i) < c_{\mathrm{u}}^{\mathcal{M}}(\tilde\pi,t_i)$ if $E_{e}(\pi,t_i)$, 
but $c_{\mathrm{u}}^{\mathcal{M}}(\tilde\pi,t_i) < c_{\mathrm{u}}^{\mathcal{M}}(\pi,t_i)$ otherwise. If end-host $e$
is entitled to use the cheaper path $\pi$, the cheaper path 
$\pi$ must thus be selected with probability $u(\pi,t|\tilde\pi) = 1$ to minimize the end-host's cost, and with probability 0 otherwise.
\end{enumerate}

In summary, for all intervals with start $t_i > t_0$, an end-host~$e$ optimizes its 
cost by switching to an alternative path $\pi$ if and only if path~$\pi$ is cheaper than
the current path $\tilde\pi$ and end-host $e$ is entitled
to use path~$\pi$. This 
path-switching behavior is exactly captured by the path-selection function
$u_{\mathrm{F}}(\pi,t|\tilde\pi)$.
Therefore, path-selection strategy $\sigma_{\mathrm{F}}$ is
an equilibrium strategy for both the initial interval and the 
subsequent intervals of the mechanism, which proves 
\cref{lem:floss:equilibrium}.

\section{The CROSS Mechanism}
\label{sec:cross}

In this section, we present a second stabilization mechanism
called CROSS (\emph{Computation-Requiring Oscillation Suppression 
System}).

\subsection{Overview}
\label{sec:cross:overview}

While the FLOSS mechanism (cf.~\cref{sec:floss}) deterministically
achieves stability at equal load, its strict enforcement of the 
migration allowance represents a problem in case of path failures.
When a path fails, an end-host on that path is not
allowed to switch to an alternative path immediately.
Only when the path failure is detected after some time 
by the mechanism, enforcement of the mechanism can be stopped
and the end-hosts can be allowed to use an alternative path. 
For highly critical transmissions, such inflexibility is 
undesirable.

The CROSS mechanism allows end-hosts to obtain
an \emph{insurance} against such cases of path failure.
Basically, the CROSS mechanism works similarly to the initial
interval of the FLOSS mechanism: End-hosts are
required to register for one path of their choice, which in general
cannot be changed during the upcoming interval. Unlike FLOSS,
however, the CROSS mechanism offers the possibility of registration 
for a second path that can be immediately used in case of a 
path failure, even if the path failure is not yet verified.

\begin{figure}
    \centering
    \includegraphics[width=\linewidth]{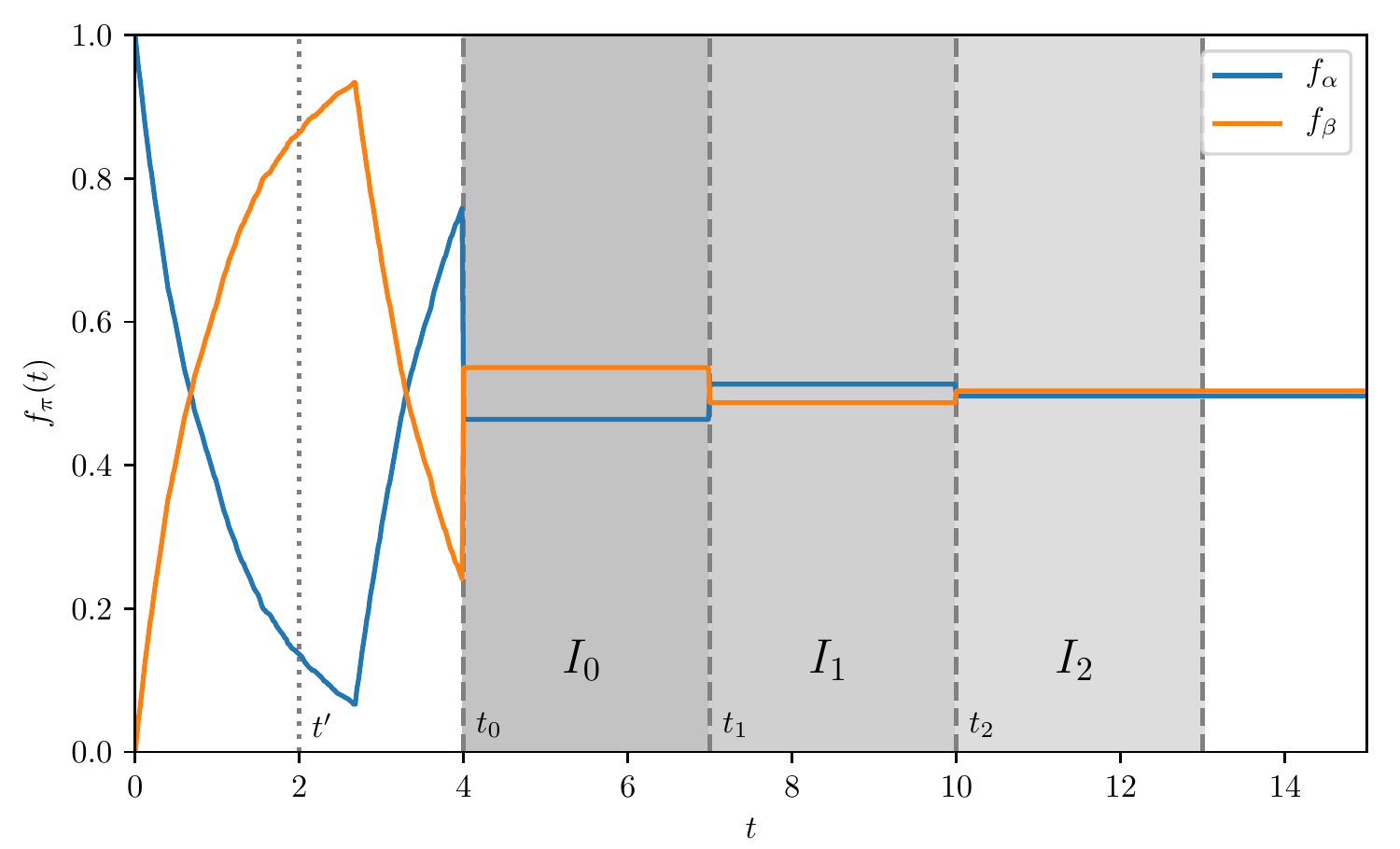}
    \vspace{-23pt}
    \caption{Simulation of CROSS enforcement in an oscillation-prone system $O = (\Pi=\{\alpha,\beta\},r=1,p=1,T=2,A_0=1,v=\{\sigma_{\mathrm{C}}\mapsto1\})$, with $\epsilon = 0.01$.}
    \label{fig:cross:simulation}
    \vspace{-15pt}
\end{figure}

However, the question is how to avoid that end-hosts
always register for both paths and, if on the more expensive path,
falsely claim to be affected by a path failure and switch to the
cheaper path. Such opportunistic behavior would cause oscillation.
To solve this problem, the idea of the CROSS mechanism is
that end-hosts must prove that they need the immediate-switching
option for insurance against path failures, 
not simply for opportunistic cost reduction.
End-hosts can prove their truthfulness by paying a price
for the immediate-switch option. This price must be
higher than any cost gain that can be achieved by switching
to a cheaper path in a scenario without path failure.
An end-host that paid this price thus only switches to
the backup path if a path failure has occurred;
if no path failure occurred, the end-host would not trade
its insurance option against the cost gain, as the insurance
option is more valuable to the end-host than any cost gain.
Immediate switching during the interval can thus be allowed to 
the end-hosts with a backup-path registration. Moreover,
immediate switching behavior by those end-hosts
is an indication of path failure, which
means that all other end-hosts must be allowed to migrate
as well.

As a price for the backup path registration, 
the CROSS mechanism requires the solution to a
computationally hard puzzle. This puzzle is structured
such that only end-hosts with a sufficiently high
valuation of the backup path will obtain a solution.
More precisely, each puzzle $\mathcal{E}$ is associated
with a cryptographic hash function $h:\{0,1\}^{\ast} 
\mapsto [0, 1]$ and a difficulty level~$\delta \geq 0$. An 
end-host~$e$ can solve a puzzle~$\mathcal{E}(\pi)$ for registering 
at a backup path~$\pi$ by 
finding a value $s$ such that $h(\pi, t_i, e, s) \leq 2^{-\delta}$,
where $t_i$ is the start of the next balancing trial. 
Given a cryptographic hash function,
a puzzle~$\mathcal{E}(\pi)$ can only be solved by brute 
force, i.e., varying~$s$ in a series of hash computations. 
By finding an appropriate~$s$, 
an end-host can obtain a backup-path registration.

Also unlike FLOSS, the CROSS mechanism allows end-hosts to 
register at a path of their choice not only for the initial
interval, but for every interval. Therefore, even if
the path failure is not detected for some reason
(e.g., because no end-host obtained a backup registration), 
the end-host can use the alternative path in the interval after a 
path failure. The CROSS mechanism thus has a
non-deterministic approach for achieving stability:
Intervals in CROSS serve as \emph{balancing trials}
and are repeated until the load imbalance is small enough
that end-hosts do not switch paths anymore.
Since the end-hosts select each path with probability~$1/2$
in any balancing trial,
the probability that an approximately equal load distribution
results after a few balancing trials is substantial.
Still, the additional flexibility of CROSS results in a loss
of convergence guarantees: Instead of convergence to an equal-load
distribution, the CROSS mechanism only guarantees convergence to
a traffic distribution with approximately equal load. A simulation of CROSS enforcement is visualized in Figure~\ref{fig:cross:simulation}, which also shows the convergence produced by
the CROSS approach.

\begin{theorem}
The CROSS mechanism is an incentive-compatible stabilization
mechanism that achieves stability at approximately
equal load, i.e., for every 
$\epsilon > 0$, $\lim_{t\rightarrow\infty} \Delta(t) < \epsilon$.
\label{thm:cross}
\end{theorem}

The CROSS mechanism achieves stability at approximately
equal load by  incentivizing the universal adoption of path-selection strategy $\sigma_{\mathrm{C}}$, 
which prescribes that end-hosts
only use a path if they have a corresponding registration
and only use a backup in case of path failures.
More formally, Theorem \ref{thm:cross} directly follows
from \cref{lem:cross:stability,lem:cross:equilibrium}:

\begin{lemma}
Universal adoption of the CROSS path-selection 
strategy~$\sigma_{\mathrm{C}}$ leads to stability at approximately
equal load.
\label{lem:cross:stability}
\end{lemma}

\begin{lemma}
Universal adoption of the CROSS path-selection 
strategy~$\sigma_{\mathrm{C}}$ represents a PSS equilibrium 
given enforcement of the CROSS mechanism.
\label{lem:cross:equilibrium}
\end{lemma}

While the proof of \cref{lem:cross:stability} is intuitive and can thus be found in~\cref{sec:cross:stability}, \cref{lem:cross:equilibrium} is proven
below.

\subsection{PSS Equilibrium Analysis}
\label{sec:cross:equilibrium}

In this section, we prove \cref{lem:cross:equilibrium}
by showing that universal adoption of
path-selection strategy $\sigma_{\mathrm{C}}$ is a PSS
equilibrium, i.e., if all other end-hosts adopt 
$\sigma_{\mathrm{C}}$, $\sigma_{\mathrm{C}}$ is the optimal
strategy for a single end-host~$e$.
The path-selection strategy $\sigma_{\mathrm{C}}$ 
is characterized by the following path-selection function 
for~$\pi~\neq~\tilde\pi$:\vspace{-5pt}\begin{equation}u_{\mathrm{C}}(\pi,t|\tilde\pi) = 
\begin{cases} 1/2 & \text{if } t = t_i \land \neg R'_{e}(\pi,t),\\ 
1 & \text{if } c_{\tilde\pi}(t-T) = \infty \land R'_{e}(\pi,t), \\
0 & \text{otherwise,} \end{cases}\end{equation}
where $t_i$ is the start time of any balancing trial, 
$c_{\tilde\pi}(t-T)~=~\infty$ designates a path failure and
$R'_e(\pi,t)$ is true if and only if end-host $e$ has
a backup registration for path $\pi$ at time $t$.
Moreover, $u_{\mathrm{C}}(\tilde\pi,t|\tilde\pi) =
1 - u_{\mathrm{C}}(\pi,t|\tilde\pi)$.

As in FLOSS, registering has cost $c_{\mathrm{a}}$,
whereas using a path without registration imposes a penalty
cost $c_{\mathrm{p}} = \infty$. 
Additionally, an end-host incurs cost by solving puzzles,
where each hashing operation has cost $c_{\mathrm{h}}$.
To an end-host with valuation~$\omega$ of a backup path,
a hash operation has the expected utility~$\mathbb{E}[U_{\mathrm{h}}](\delta,\omega) = 
2^{-\delta}\omega-c_{\mathrm{h}}$.

Given puzzle-difficulty level~$\delta$, an end-host
thus solves a puzzle if and only if it has a
backup valuation $\omega$ such that 
$\mathbb{E}[U_{\mathrm{h}}](\delta,\omega) > 0$.
% This puzzle-solving cost is structured as follows.  The total cost
% of puzzle solving
% for an end-host depends on the number of performed 
% hash computations, which is bounded by the end-host's valuation $v$ 
% of a backup path: An end-host $e$ with a backup valuation $v$ will
% perform at most $n_{\mathrm{max}}(v)= \lfloor
% (v-c_{\mathrm{a}})/c_{\mathrm{h}}\rfloor$ hash computations to solve
% the puzzle and thus find a solution with probability 
% $p_{\mathcal{E}}(\delta, v) = 
% 1-(1-2^{-\delta})^{n_{\mathrm{max}}(v)}$.
If an end-host does not solve a puzzle, it simply obtains
a regular registration for one path at cost $c_{\mathrm{a}}$, 
where every path is selected with probability 1/2. 
Obtaining no registration and using any path would incur a much
higher penalty cost $c_{\mathrm{p}} \gg c_{\mathrm{a}}$ and
is thus not rational. Therefore, an end-host with a
registration for one path uses this path from the start~$t_i$
of the balancing trial. If an end-host solves
a puzzle, the end-host obtains a backup registration for the
path corresponding to the puzzle and obtains a regular registration
for the other path at cost $c_{\mathrm{a}}$. 
Since CROSS enforces that an end-host
can only switch once to its backup path and never switch back
during the balancing trial, every end-host with a backup-path
registration starts by using the path with its regular registration 
at time $t_i$. In summary, the optimal path-selection function for all $t = t_i$ is $u_{\mathrm{C}}(\pi,t|\pi') = 
1/2$ if $\neg R_e'(\pi,t)$.

During the balancing trial, no reallocation decisions are taken 
before $t_i + T$, as the expected path costs during
$[t_i, t_i + T]$ is $\mathbb{E}[c_{\alpha}] = \mathbb{E}[c_{\beta}]
= 1/2^p$. Only at $t_i + T$, the actual imbalance $\Delta(t) =
|f_{\pi}(t_i) - f_{\tilde\pi}(t_i)|$ between a more expensive path~$\tilde\pi$
and a cheaper path $\pi$ becomes visible to the end-hosts.
If the end-hosts on path $\tilde\pi$ with a backup registration
for path~$\pi$ switched at that point, they would  
save $\Delta \overline{C} =
\int_{t_i+T}^{t_{i+1}} (c_{\tilde\pi}(t) - c_{\pi}(t))\, \mathrm{d}t$,
which is bounded above by $\Delta \overline{C}_{\mathrm{max}} = t_{i+1}-t_i-T$.
However, such a switch would erase the backup value~$\omega$ of 
path~$\pi$ for the end-host, which is why an end-host with backup 
registration for path~$\pi$ only switches to path~$\pi$
if $\omega < \Delta\overline{C}$. In order to disincentivize such migration
and keep the load distribution constant, the CROSS mechanism
chooses the puzzle-difficulty level~$\delta$ such 
that~$\mathbb{E}[U_{\mathrm{h}}](\delta,v) > 0$
if and only if $\omega > \Delta \overline{C}_{\mathrm{max}}$. 
This choice of~$\delta$ leads to a situation
where the end-hosts with a backup registration will only
switch to the backup path in case of a path failure,
as these end-hosts value the backup option higher
than any cost reduction obtainable without path failure.
In case of a path failure, however, trading the backup value~$\omega$
of path $\pi$ against the infinite cost of failed path~$\tilde\pi$
is rational and the end-hosts with a backup registration switch
the paths. In summary, the optimal path-selection function
for end-host $e$ and for all $t \neq t_i$ 
is thus $u_{\mathrm{C}}(\pi,t|\tilde\pi) = 1$ if $R_e'(\pi,t)$ and 
$c_{\tilde\pi}(t-T) = \infty$,  
and $u_{\mathrm{C}}(\pi,t|\tilde\pi) = 0$ otherwise.
Thereby, path-selection strategy $\sigma_C$ has been established
as the PSS equilibrium strategy.
\section{Practical Application}
\label{sec:practical}

While the focus of this paper is on the theoretical exploration
of selfish path selection and stabilization mechanisms, this section lays out a pathway toward practical application of our findings.
First, we discuss practical requirements for inter-domain 
stabilization mechanisms in \cref{sec:practical:requirements}. 
In \cref{sec:practical:enforcement-architecture}, we present a 
mechanism-enforcement architecture that conforms to these requirements. 
In \cref{sec:practical:floss} and 
\cref{sec:practical:cross}, we outline how the FLOSS
and CROSS mechanisms could be practically implemented.

\subsection{Requirements}
\label{sec:practical:requirements}

If a stabilization mechanism is to be practically applied by
network operators in an inter-domain architecture, the mechanism must
conform to the following requirements:

\begin{enumerate}
    \item \textit{Limited overhead:} The stabilization mechanism 
    must only induce a small overhead on the systems of network 
    operators. In particular, the genuine function of AS border 
    routers (forwarding traffic at line rate) must not be 
    compromised by expensive mechanism-enforcement tasks. Note that both mechanisms only need to be enforced by routers in case of oscillation and until
    stabilization is achieved; however, the mechanisms should induce little overhead even during this short time span.
    \item \textit{No explicit inter-AS coordination 
    (coordination-freeness):} The stabilization mechanism must not 
    rely on explicit inter-AS coordination. Such explicit 
    coordination may not be feasible or scalable, as the domains 
    that perceive the same oscillation pattern may be mutually 
    unknown, mutually distrusted, or very distant from each other.
\end{enumerate}

\subsection{Mechanism-Enforcement Architecture}
\label{sec:practical:enforcement-architecture}

To enforce a stabilization mechanism, an AS operator needs
the means to detect, inform, and punish the selfish entities 
that employ an oscillatory path-selection strategy. In this section,
we describe a mechanism-enforcement architecture that provides these
means to an AS operator while conforming to the requirements in \cref{sec:practical:requirements}.

From an inter-domain perspective, the most important architectural 
question is the question of \emph{coordination}, i.e., how each AS 
perceiving an oscillation pattern contributes to oscillation 
suppression. As explicit inter-AS coordination is undesirable, an 
implicit method for \textit{responsibility assignment} is necessary. 
\begin{figure}
    \centering
    \begin{subfigure}[b]{0.22\linewidth}
    \centering
        \begin{tikzpicture}[
	asnode/.style={circle, draw=black!60, shading=radial,outer color={rgb,255:red,137;green,207;blue,240},inner color=white, thick, minimum size=8mm},
	hostnode/.style={circle, draw=black!60, shading=radial,outer color={rgb,255:red,255;green,153;blue,102},inner color=white, thick, minimum size=5mm},
	oscnode/.style={circle, draw=black, ultra thick, minimum size=10mm},
	]
	\node[hostnode]	 (O)	  at (0,5)	{\textbf{O}}; 
	\node[hostnode]	 (D)	  at (0,1)	{\textbf{D}};
	
	\node[asnode]	 (asA1)	  at (0,4)	{$\boldsymbol{A_1}$}; 
	\node[asnode]	 (asA4)	  at (0,2)	{$\boldsymbol{A_4}$};
	\node[oscnode]   at (asA1) {};
	
	\draw[-] (O.south) to (asA1.north);
	\draw[-] (asA4.south) to (D.north);
	
	\draw[-] (asA1.south west) to [out=250,in=110] (asA4.north west);
	\draw[-] (asA1.south east) to [out=290,in=70] (asA4.north east);
	
	\node[circle,fill=black,thick] at (asA1.south west) {};
	\node[circle,fill=black,thick] at (asA1.south east) {};
	
	\node[asnode]	 (asA2)	  at (-0.5,3)	{$\boldsymbol{A_2}$}; 
	\node[asnode]	 (asA3)	  at (0.5,3)	{$\boldsymbol{A_3}$};
	
	\draw[-,red,ultra thick] (O.south)  to[spline through={(asA1.west)(asA1.south west)(asA2.west)(asA2.290)(asA4.west)}] (D.north);
	\draw[-,blue,ultra thick] (O.south) to[spline through={(asA1.east)(asA1.south east)(asA3.east)(asA3.250)(asA4.east)}] (D.north);
	
	\node[fill=red, minimum width=2mm, minimum height=2mm] (redSquare) at (-0.7,6.7) {};
	\node[red] at (-0.3,6.7) {$\pi_1$};
	\node[fill=blue, minimum width=2mm, minimum height=2mm] (blueSquare) at (0.3,6.7) {};
	\node[blue] at (0.7,6.7) {$\pi_2$};
	\node[fill=orange, minimum width=2mm, minimum height=2mm] (orangeSquare) at (-0.7,6.3) {};
	\node[orange] at (-0.3,6.3) {$\pi_3$};
	\node[fill=violet, minimum width=2mm, minimum height=2mm] (violetSquare) at (0.3,6.3) {};
	\node[violet] at (0.7,6.3) {$\pi_4$};
	
\end{tikzpicture}
        \caption{}
        \label{fig:mechanisms:enforcement-architecture:type-1}
    \end{subfigure}
    \begin{subfigure}[b]{0.22\linewidth}
        \centering
        \begin{tikzpicture}[
	asnode/.style={circle, draw=black!60, shading=radial,outer color={rgb,255:red,137;green,207;blue,240},inner color=white, thick, minimum size=8mm},
	hostnode/.style={circle, draw=black!60, shading=radial,outer color={rgb,255:red,255;green,153;blue,102},inner color=white, thick, minimum size=5mm},
	oscnode/.style={circle, draw=black, ultra thick, minimum size=10mm},
	]
	\node[hostnode]	 (O)	  at (0,6)	{\textbf{O}}; 
	\node[hostnode]	 (D)	  at (0,1)	{\textbf{D}};
	
	\node[asnode]    (asA0)   at (0,5)  {$\boldsymbol{A_0}$};
	\node[asnode]	 (asA1)	  at (0,4)	{$\boldsymbol{A_1}$}; 
	\node[asnode]	 (asA4)	  at (0,2)	{$\boldsymbol{A_4}$};
	\node[oscnode]   at (asA1) {};
	
	\draw[-] (O.south) to (asA0.north);
	\draw[-] (asA0.south) to (asA1.north);
	\draw[-] (asA4.south) to (D.north);
	
	\draw[-] (asA1.south west) to [out=250,in=110] (asA4.north west);
	\draw[-] (asA1.south east) to [out=290,in=70] (asA4.north east);
	
	\node[circle,fill=black,thick] at (asA1.south west) {};
	\node[circle,fill=black,thick] at (asA1.south east) {};
	
	\node[asnode]	 (asA2)	  at (-0.5,3)	{$\boldsymbol{A_2}$}; 
	\node[asnode]	 (asA3)	  at (0.5,3)	{$\boldsymbol{A_3}$};
	
	\draw[-,red,ultra thick] (O.south)  to[spline through={(asA0.west)(asA0.south)(asA1.west)(asA1.south west)(asA2.west)(asA2.290)(asA4.west)}] (D.north);
	\draw[-,blue,ultra thick] (O.south) to[spline through={(asA0.east)(asA0.south)(asA1.east)(asA1.south east)(asA3.east)(asA3.250)(asA4.east)}] (D.north);
	
\end{tikzpicture}
       \caption{}
        \label{fig:mechanisms:enforcement-architecture:type-2}
    \end{subfigure}
    \begin{subfigure}[b]{0.25\linewidth}
        \centering
        \begin{tikzpicture}[
	asnode/.style={circle, draw=black!60, shading=radial,outer color={rgb,255:red,137;green,207;blue,240},inner color=white, thick, minimum size=8mm},
	hostnode/.style={circle, draw=black!60, shading=radial,outer color={rgb,255:red,255;green,153;blue,102},inner color=white, thick, minimum size=5mm},
	oscnode/.style={circle, draw=black, ultra thick, minimum size=10mm},
	]
	\node[hostnode]	 (O)	  at (0,6)	{\textbf{O}}; 
	\node[hostnode]	 (D)	  at (0,1)	{\textbf{D}};
	
	\node[asnode]    (asA0)   at (0,5)  {$\boldsymbol{A_0}$};
	\node[asnode]	 (asA1)	  at (0,4)	{$\boldsymbol{A_1}$}; 
	\node[asnode]	 (asA4)	  at (0,2)	{$\boldsymbol{A_4}$};
	\node[oscnode]   at (asA1) {};
	\node[oscnode]   at (asA0) {};
	
	\draw[-] (O.south) to (asA0.north);
	\draw[-] (asA0.south) to (asA1.north);
	\draw[-] (asA4.south) to (D.north);
	\draw[-] (asA0.east) to [out=320, in=40] (asA4.east);
	
	\draw[-] (asA1.south west) to [out=250,in=110] (asA4.north west);
	\draw[-] (asA1.south east) to [out=290,in=70] (asA4.north east);
	
	\node[circle,fill=black,thick] at (asA1.south west) {};
	\node[circle,fill=black,thick] at (asA1.south east) {};
	\node[circle,fill=black,thick] at (asA0.south) {};
	\node[circle,fill=black,thick] at (asA0.east) {};
	
	\node[asnode]	 (asA2)	  at (-0.5,3)	{$\boldsymbol{A_2}$}; 
	\node[asnode]	 (asA3)	  at (0.5,3)	{$\boldsymbol{A_3}$};
	
	\draw[-,red,ultra thick] (O.south)  to[spline through={(asA0.west)(asA0.south)(asA1.west)(asA1.south west)(asA2.west)(asA2.290)(asA4.west)}] (D.north);
	\draw[-,blue,ultra thick] (O.south) to[spline through={(asA0.west)(asA0.south)(asA1.east)(asA1.south east)(asA3.east)(asA3.250)(asA4.east)}] (D.north);
	\draw[-,orange,ultra thick] (O.south) to[spline through={(asA0.east)(1.2,3.5)(asA4.east)}] (D.north);
	
\end{tikzpicture}
       \caption{}
        \label{fig:mechanisms:enforcement-architecture:type-3}
    \end{subfigure}
    \begin{subfigure}[b]{0.25\linewidth}
        \centering
       \begin{tikzpicture}[
asnode/.style={circle, draw=black!60, shading=radial,outer color={rgb,255:red,137;green,207;blue,240},inner color=white, thick, minimum size=8mm},
hostnode/.style={circle, draw=black!60, shading=radial,outer color={rgb,255:red,255;green,153;blue,102},inner color=white, thick, minimum size=5mm},
oscnode/.style={circle, draw=black, ultra thick, minimum size=10mm},
]
\node[hostnode]	 (O)	  at (0,7)	{\textbf{O}}; 
\node[hostnode]	 (D)	  at (0,1)	{\textbf{D}};

\node[asnode]	 (asA1)	  at (0,6)	{$\boldsymbol{A_1}$}; 
\node[asnode]	 (asA4)	  at (0,4)	{$\boldsymbol{A_4}$};
\node[asnode]	 (asA7)	  at (0,2)	{$\boldsymbol{A_7}$};
\node[oscnode]   at (asA1) {};
\node[oscnode]   at (asA4) {};

\draw[-] (O.south) to (asA1.north);
\draw[-] (asA7.south) to (D.north);

\draw[-] (asA1.south west) to [out=250,in=110] (asA4.north west);
\draw[-] (asA1.south east) to [out=290,in=70]  (asA4.north east);
\draw[-] (asA4.south west) to [out=250,in=110] (asA7.north west);
\draw[-] (asA4.south east) to [out=290,in=70]  (asA7.north east);

\node[circle,fill=black,thick] at (asA1.south west) {};
\node[circle,fill=black,thick] at (asA1.south east) {};
\node[circle,fill=black,thick] at (asA4.south west) {};
\node[circle,fill=black,thick] at (asA4.south east) {};

\node[asnode]	 (asA2)	  at (-0.5, 5)	{$\boldsymbol{A_2}$}; 
\node[asnode]	 (asA3)	  at (0.5,  5)	{$\boldsymbol{A_3}$};
\node[asnode]	 (asA5)	  at (-0.5, 3)	{$\boldsymbol{A_5}$};
\node[asnode]	 (asA6)	  at (0.5, 3)	{$\boldsymbol{A_6}$};

\draw[-,red,ultra thick] (O.south) to[spline through={(asA1.west)(asA1.south west)(asA2.70)(asA2.west)(asA2.290)(asA4.north west)(asA4.south west)(asA5.north)(asA5.west)(asA5.south)(asA7.north west)(asA7.west)}] (D.north);
\draw[-,blue,ultra thick] (O.south) to[spline through={(asA1.west)(asA1.south west)(asA2.70)(asA2.east)(asA2.290)(asA4.north west)(asA4.east)(asA4.south east)(asA6.110)(asA6.west)(asA6.250)(asA7.east)}] (D.north);
\draw[-,orange,ultra thick] (O.south) to[spline through={(asA1.east)(asA1.south east)(asA3.110)(asA3.west)(asA3.250)(asA4.north east)(asA4.west)(asA4.south west)(asA5.70)(asA5.east)(asA5.290)(asA7.west)}] (D.north);
\draw[-,violet,ultra thick] (O.south) to[spline through={(asA1.east)(asA1.south east)(asA3.110)(asA3.east)(asA3.250)(asA4.north east)(asA4.south east)(asA6.north)(asA6.east)(asA6.south)(asA7.north east)(asA7.east)}] (D.north);

\node[] at (0,-0.1) {};
\end{tikzpicture}
       \vspace{-25pt}
       \caption{}
        \label{fig:mechanisms:enforcement-architecture:type-4}
    \end{subfigure}
    \vspace{-10pt}
    \caption{Oscillation patterns.}
    \label{fig:mechanisms:patterns}
    \vspace{-15pt}
\end{figure}
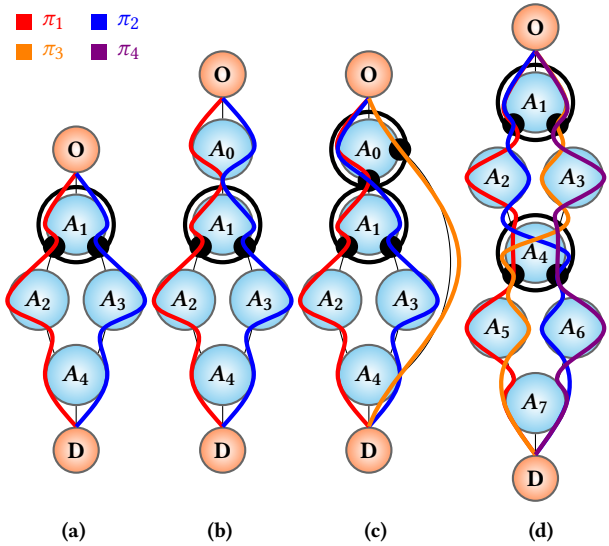

We leverage a fundamental property of paths in inter-domain network 
graphs as a natural way to assign responsibility for inter-domain 
oscillation suppression. This fundamental property is based on the 
following insight: For every pair of paths connecting the same 
origin and destination ASes, there is at least one AS (henceforth: 
the \emph{splitting AS}) in which the paths split, i.e., 
the paths contain different egress interfaces out of the AS. 
For every oscillation between two paths, there is thus at least one 
AS which perceives the oscillation as an oscillation of traffic 
between egress interfaces, not only as periodic upswings and 
downswings in the load at one egress interface. 
Such splitting ASes are the natural candidates for 
a leading role in inter-domain oscillation suppression, 
as these ASes are both best informed about the oscillation and 
in the best position to manage the oscillating traffic.

For illustration of the path-splitting property, 
\cref{fig:mechanisms:patterns} shows different types of 
oscillation patterns for paths connecting an origin end-host $O$ and
a destination end-host $D$. In the simplest cases, the oscillation 
may be perceived at the origin AS (AS $A_1$ in 
\cref{fig:mechanisms:enforcement-architecture:type-1}) or at one 
intermediate AS (AS $A_1$ in 
\cref{fig:mechanisms:enforcement-architecture:type-2}). However, 
the oscillation may be perceived at multiple 
splitting ASes. The different paths may pass through a different 
number of egress interfaces at which the mechanism is enforced. For 
example, path $\pi_3$ in 
\cref{fig:mechanisms:enforcement-architecture:type-3} only passes 
through one critical egress interface (at AS $A_0$), whereas paths 
$\pi_1$ and~$\pi_2$ pass through two critical egress 
interfaces. Conversely, each path in 
\cref{fig:mechanisms:enforcement-architecture:type-4} passes through
two egress interfaces at which a load-balancing mechanism is 
enforced. Any stabilization mechanism may thus be applied
repeatedly and with different frequency to flows belonging to the 
same oscillation-prone system.

\begin{figure}
    \centering
    \begin{tikzpicture}[
asnode/.style={circle, draw=black!60, shading=radial,outer color={rgb,255:red,137;green,207;blue,240},inner color=white, thick, minimum size=60mm},
routernode/.style={circle, draw=black!60, shading=radial,outer color={rgb,255:red,102;green,200;blue,102},inner color=white, thick, minimum size=5mm},
oscserv/.style={draw=black!50, shading=radial, outer color={rgb,255:red,200;green,100;blue,200},inner color=white, thick, minimum size=10mm},
statsquare/.style={rectangle,draw=black!60, fill=white, minimum width=10mm,minimum height=8mm},
puzzlesquare/.style={draw=black!50, shading=radial, outer color={rgb,255:red,200;green,200;blue,200},inner color=white, thick, minimum width=10mm},
]

\node[asnode]	 (asA1)	  at (0,0)	{};

\node[routernode] (r3) at (asA1.north east) {$r_3$};
\node[routernode] (r4) at (asA1.east) {$r_4$}; 
\node[routernode] (r5) at (asA1.south east) {$r_5$};

\node[routernode] (r1) at (asA1.north west) {$r_1$};
\node[routernode] (r2) at (asA1.south west) {$r_2$};

\draw[-] (r1.north west) + (-0.5,+0.5) -- (r1.north west);
\draw[-] (r4.east) + (1,0) -- (r4.east);
\draw[-] (r3.north east) + (+0.5,+0.5) -- (r3.north east);
\draw[-] (r2.south west) + (-0.5,-0.5) -- (r2.south west);
\draw[-] (r5.south east) + (0.5,-0.5) -- (r5.south east);

%\draw[->,red,ultra thick] (r1.north west) + (-1,+1) to[spline through={(r1.north)(r3.north)}] (3.5, 3.5);
%\node[red] at (3.6,3) {$\pi_3$};
%\draw[->,blue,ultra thick] (r1.north west) + (-1,+1) to[spline through={(r1.south)(r4.north)}] (4.5, 0);
%\node[blue] at (4,0.3) {$\pi_4$};
%\draw[->,dashed,orange,ultra thick] (r1.north west) + (-1,+1) to[spline through={(r1.south)(r5.north)}] (3.5, -3.5);
%\node[orange] at (3.5,-2.7) {$\pi_{51}$};
%\draw[->,dashed,brown,ultra thick] (r2.south west) + (-1,-1) to[spline through={(r2.south)(r5.south)}] (3.5, -3.5);
%\node[brown] at (2.3,-3) {$\pi_{52}$};

\node[oscserv] (os) at (-1,-1) {\textbf{OSS}};

\draw[->,dotted,thick] (r3.south west) -- (os.north east);
\draw[->,dotted,thick] (r4.west) -- (os.east);
\draw[->,dotted,thick] (r5.north west) -- (os.south east);

\node[statsquare] at(1.5,1.3) {};
\node at (1.3,1.1) {stat};
\draw[-,thick,gray] (1.8, 1.3) sin (1.65, 1.1) cos (1.5, 1.3) sin (1.35, 1.5) cos (1.2,1.3);

\node[statsquare] at(1.8,-0.5) {};
\node at (1.95,-0.75) {stat};
\draw[-,thick,gray] (2.1,-0.5) sin (1.95,-0.3) cos (1.8,-0.5) sin (1.65,-0.7) cos (1.5,-0.5);

\node[statsquare] at(1.2,-1.8) {};
\node at (1.35,-1.55) {stat};
\draw[-,thick,gray] (1.5,-1.8) -- (0.9,-1.8);

\draw[->,dashed,thick] (os.north west) .. controls +(-1,1) .. node[puzzlesquare] {$M_3$} (r3.west);
\draw[->,dashed,thick] (os.north) .. controls +(0,1) .. node[puzzlesquare] {$M_4$} (r4.west);

\coordinate (r1up) at ($ (r1.north west) + (-0.5,+0.5)$);
\coordinate (r3up) at ($ (r3.north east) + (+0.5,+0.5) $);
\coordinate (r4up) at ($ (r4.east) + (+1,+0) $);

\draw[-,red,ultra thick] (r1up)  to[spline through={(r1.north)(r3.north)}] (r3up);
\draw[-,blue,ultra thick] (r1up)  to[spline through={(r1.south)(r4.north)}] (r4up);
\node[red] at (0,2.1) {$\pi_1$};
\node[blue] at (0,0.9) {$\pi_2$};

\end{tikzpicture}
    \vspace{-10pt}
    \caption{Mechanism-enforcement architecture (within the splitting AS).}
    \label{fig:mechanisms:enforcement-architecture:architecture}
    \vspace{-15pt}
\end{figure}
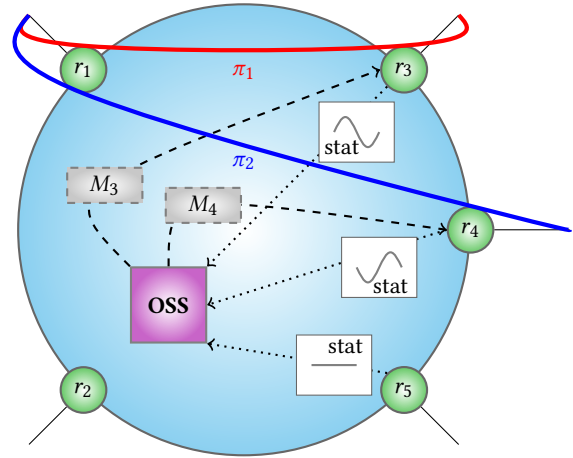

In the intra-domain context, the mechanism-enforcement 
architecture envisages a centralized oscillation-suppression 
service  (OSS) in each AS. 
The OSS is capable of interacting with the border 
routers at the egress interfaces. For a
splitting AS, this OSS functions as displayed in 
\cref{fig:mechanisms:enforcement-architecture:architecture}. By 
collecting aggregate load statistics from the border routers, the OSS in the splitting AS can identify the egress interfaces
between which oscillation occurs (through correlation). As the 
presence of such oscillation means that the AS is obliged to 
enforce a stabilization mechanism, the OSS equips every 
oscillation-perceiving border router $r_i$ with data $M_i$ that is 
necessary to enforce the mechanism (e.g., start time of the next interval). By further collecting load 
statistics from the egresses, the OSS monitors and 
continuously adapts the execution of the mechanism.
The border routers communicate with the origins of the oscillating
flows by appending mechanism-relevant information
to passing packets.

\subsection{FLOSS in Practice}
\label{sec:practical:floss}

In the following, we discuss how the FLOSS mechanism
could be applied by the mechanism-enforcement architecture
from \cref{sec:practical:enforcement-architecture}, while
conforming to the practicality requirements laid out in
\cref{sec:practical:requirements}, namely limited overhead
and coordination-freeness.

\subsubsection{Limited Overhead}
\label{sec:practical:floss:overhead}

\paragraph*{Registration on routers} In order to signal that
end-hosts must register for an upcoming time interval, 
a border router appends the start time $t_i$ of the next interval to
passing packets. If an end-host witnesses such a call for 
registrations in its packets, it can send a packet with 
a registration request over the desired egress.
A border router can keep track of registrations using a Bloom 
filter, which approximates a set of flow IDs. A Bloom filter 
offers  constant complexity for both lookup and insertion, 
although suffering from false positives. 
When checking for registrations, false positives result in 
unregistered flows being able to send over an egress and being 
rewarded like loyal flows. 
However, the enforced migration rate $\rho$ can simply be 
discounted by the false-positive rate of the Bloom filter 
such that the desired migration rate is enforced despite the 
presence of lucky unregistered flows.

\paragraph*{Enforcement of single registration} In order to avoid 
that an end-host registers on multiple egresses, a border router 
forwards all registrations to the OSS, which keeps track of 
egress-specific registration by flows and can therefore spot 
multiple registrations by the same flow. If multiple registrations 
are detected, the OSS pushes a blacklist update for the malicious 
flow ID to the border routers. In order to avoid introducing DoS 
attacks where a malicious actor provokes the blacklisting of an 
end-host by sending multiple registrations, we assume some form of 
lightweight source authentication, which is typically offered by 
path-aware Internet architectures \cite{rot2020piskes}.

\paragraph*{Selective admission of migrating flows} 
Border routers need an efficient way to decide whether to grant registration applications to flows that are willing to switch
paths, while preserving the property that a maximum share 
$\rho$ of flows migrates. Such selective admission can be 
implemented using a publicly know hash function $h$, which maps the 
flow ID $f$ to the interval $[0,1]$. If $h(t_i|f) < \rho$, the 
registration is granted, where $t_i$ is the beginning time of the 
next registration-enforcement interval. This construction has the 
advantage that an end-host can locally check whether it will be 
accepted on the alternative ingress, as $h$, $t_i$, and $f$ are 
known to the end-host. Therefore, the border router is not bothered 
by registration requests from end-hosts that would be rejected. 
Furthermore, it is important to choose the flow ID~$f$ based on 
attributes that the source end-host cannot easily influence without 
compromising its communication, e.g., source and destination IP, 
but not source or destination port.

\paragraph*{Small traffic allowance for unregistered flows} While 
unregistered end-hosts should not be able to properly use an egress,
these end-hosts should be able to send a few packets over the egress
to measure the latency of the corresponding path. Also, short flows,
e.g., DNS requests, should not be required to 
obtain a registration. Such a limited 
traffic allowance can be efficiently achieved by applying the 
mechanism only to a subset of packets, e.g., by sub-sampling. If 
registrations are only checked for a sub-set of packets, even an 
unregistered flow has a high chance of getting a few packets through
the egress, while still experiencing severe disruption when sending 
a large number of packets over the egress. Due to the structure
of congestion-control algorithms, sub-sampling rates as low as 1\% 
already cause enough packet drops to make a path completely 
unusable for unregistered flows~\cite{lukaseder2016comparison}.
Moreover, sub-sampling reduces the workload on border routers.

\paragraph*{Addition of new flows} In reality, new flows appear 
during the execution of the mechanism. Clearly,
these flows cannot register in advance for an enforcement
interval, as these flows do not exist beforehand. Therefore,
new flows are also allowed to register at one path of their choice
\emph{during} an enforcement interval. In order to distinguish new 
flows from flows that merely pretend to be new, the FLOSS mechanism
samples the active flows at both egresses in every interval
and inserts them into a Bloom filter. These previously active flows
are supposed to have a registration in the subsequent interval.
In contrast, truly new flows can be identified with a 
lookup failure in the mentioned Bloom filter. Due to false 
positives, a truly new flow might be mistaken for a previously 
active flow and thus be denied a retroactive registration.
However, given a small false-positive probability, the probability
that such a mistake appears at multiple egresses is negligible
such that registration at one path should always be possible
in practice. As all new flows (except the false-positive
new flows) during an interval must be expected to flock to the 
cheaper path, the migration allowance must be discounted by the 
birth rate of flows. 

\subsubsection{Coordination-Freeness}
\label{sec:practical:floss:coordination}

If there is one splitting AS for an oscillation-prone
system, there are no unintended effects
due to distributed application of the mechanism. 
However, as explained 
in \cref{sec:practical:enforcement-architecture},
there may be multiple mechanism-enforcing ASes along a path.
If $n_i$ is the number of splitting ASes along path $\pi_i$,
the costs for obtaining a registration for~$\pi_i$ and
for using~$\pi_i$ without a registration 
are~$n_i\cdot c_{\mathrm{a}}$ and~$n_i\cdot c_{\mathrm{p}}$,
respectively. In cases where $n_i$ is the same for every path $\pi_i$ of an oscillation pattern
(such as in Figure~\ref{fig:mechanisms:enforcement-architecture:type-4}),
the incentives for the end-hosts thus do not change
compared to a single-application scenario. 
However, if $n_i$ is different for the paths $\pi_i$
in the oscillation-prone system (such as in
Figure~\ref{fig:mechanisms:enforcement-architecture:type-3}),
the registration cost for
different paths may be different. For example, the registration
cost for obtaining a registration of path~$\pi_3$ in 
Figure~\ref{fig:mechanisms:enforcement-architecture:type-3}
is $c_{\mathrm{a}}$, whereas the corresponding cost
for paths~$\pi_1$ and~$\pi_2$ is~$2c_{\mathrm{a}}$. 
Since $c_{\mathrm{p}} = \infty > n_i c_{\mathrm{a}}$ for all
finite $n_i$, registering for a path is still worthwhile. 
However, an equilibrium between the two
egresses of AS $A_0$ is only reached if 
$(f_{\pi_1}+f_{\pi_2})^p + 2c_{\mathrm{a}} =
f_{\pi_3}^p + c_{\mathrm{a}}$, which implies stability
at \emph{unequal} load. However, since the 
cost~$c_{\mathrm{a}}$ for obtaining a registration is
modest (just a single packet as explained in 
\cref{sec:practical:floss:overhead}), the resulting 
load imbalance between the ASes is also modest.
Therefore, no explicit inter-AS coordination is needed.

\subsection{CROSS in Practice}
\label{sec:practical:cross}

In this section, we discuss the CROSS mechanism with
respect to the two practicality requirements.

\subsubsection{Limited Overhead}
\label{sec:practical:cross:overhead}

Compared to FLOSS, the only 
additional piece of functionality needed for CROSS is
puzzle verification. Efficient puzzle-solution verification
on border routers is performed by a hash function evaluation with the appropriate arguments, among which is the solution value provided by the data packet (cf.~\cref{sec:cross:equilibrium}).

\subsubsection{Coordination-Freeness}
\label{sec:practical:cross:coordination}

Like FLOSS, CROSS suffers from the minor issue that some paths may
require more registrations than other paths. Concerning
backup registrations, multiple applications of the mechanism
do not constitute a problem, as an end-host always has to
solve only one puzzle to obtain a backup registration. 
For example, an end-host in the network of 
Figure~\ref{fig:mechanisms:enforcement-architecture:type-3}
could insure against path failure as follows. At AS $A_0$,
the end-host would obtain a normal registration for~$\pi_3$
and a backup registration for~$\pi_1$ \emph{and}~$\pi_2$.
Such a combined backup registration is possible
by including only the respective egress of AS~$A_0$ in the
puzzle solution, not the specific path.
At AS $A_1$, the end-host can then obtain a normal registration
for one of these paths, e.g.,~$\pi_1$. 
If the end-host desires an additional insurance against failure of 
path~$\pi_1$, the end-host can solve a puzzle to obtain a backup 
registration for~$\pi_2$ at AS $A_1$. Since only one puzzle
per backup path is needed, no explicit inter-AS coordination
is necessary to preserve the incentives
of the CROSS mechanism. 

\section{Related Work}
\label{sec:related-work}

Prior research has devised traffic-engineering tools to improve network stability. However, due to the traditional paradigm of network-controlled path selection, most tools assume that packet forwarding is performed by series of decisions taken by the hops along a path. Systems such as AMP~\cite{gojmerac2003adaptive}, ReplEx~\cite{fischer2006replex}, Homeostasis~\cite{kvalbein2009multipath}, and HALO~\cite{michael2014halo} thus prescribe how routers along a path should take forwarding decisions, mostly by adapting traffic-splitting ratios based on network information. If packets must be forwarded along a path chosen by the end-host, these schemes cannot be used.

An alternative line of work is generally compatible with the emerging paradigm of end-point path selection. Assuming source routing, this flavor of research prescribes path-selection strategies that lead to convergence. However, such convergent path-selection strategies are always designed for an intra-domain context, i.e., for path selection within a domain where end-points are under control of the network operator. Due to the selfishness of end-hosts in the inter-domain context, these schemes are thus impractical. For example, Proportional Sticky Routing \cite{nelakuditi2002adaptive} relies on self-restraint of end-points, which leads to persistent preference of shortest paths over alternative paths even when alternative paths are more attractive. The convergence of MATE \cite{elwalid2002mate} and the rerouting strategy designed by Kelly and Voice \cite{kelly2005stability} is built on the assumption that the end-points restrain themselves to a maximum speed when reallocating traffic on cheaper paths, which cannot be expected from selfish end-hosts. In TeXCP \cite{kandula2005walking}, end-points are expected to comply with maximum traffic-reallocation allowances dynamically set by the network. Similarly, the rerouting policies designed by Fischer and V\"ocking \cite{fischer2009adaptive} require that end-hosts do not exceed a certain probability for switching to a cheaper path. Finally, OPS \cite{jonglez2017distributed} also demands behavior from end-hosts that is irrational in a game-theoretic sense, in particular the probabilistic usage of sub-optimal paths.

Inter-domain traffic engineering by means of incentives has only been studied in context of the BGP ecosystem, thus not accounting for path choice by end-hosts. Given rational ASes, there are different methods to achieve stability for inter-domain traffic: incentive-compatible yet oscillation-free BGP policies \cite{yang2005route, feigenbaum2006incentive}, egress-router selection under QoS constraints \cite{ho2004incentive}, cooperative traffic-engineering agreements between ASes reached by Nash bargaining \cite{shrimali2009cooperative}, and the use of prices as traffic-steering incentives~\cite{mortier2003incentive}.
\section{Conclusion}
\label{sec:conclusion}

In this work, we have set up a game-theoretic framework that allows to 
test path-selection strategies on their viability for selfish 
end-hosts, i.e., to show whether it is rational for an end-host to 
adopt a path-selection strategy, given that all other end-hosts use 
said path-selection strategy. Only strategies that form such 
equilibria may be adopted in an Internet environment,
where end-hosts are self-interested and uncontrolled.

Using this framework, we have shown that the non-oscillatory 
path-selection strategies traditionally proposed in the literature are 
not rational strategies and thus cannot be expected to be adopted by 
selfish, unrestricted end-hosts. This insight suggests that end-hosts 
must be incentivized to abstain from oscillatory path selection by 
means of stabilization mechanisms. We have designed two stabilization 
mechanisms and proved their incentive compatibility.

We understand our work as a first step and we believe that it opens 
several interesting avenues for future research. In particular, it 
would be interesting to quantify the cost
of oscillation to a network and to investigate its relationship to 
the network type. Comparing the oscillation cost to
the overhead of stabilization mechanisms
would then allow to characterize the conditions under which the 
employment of stabilization mechanisms is appropriate.

% Finally, an interesting question is: How plausible is the rationality assumption from game theory in the Internet? After all, ubiquitous protocols such as TCP~\cite{akella2002selfish,godfrey2010incentive} and BitTorrent~\cite{locher2006freeriding} are widely known to allow abuse by selfish agents, but still most users faithfully adhere to these protocols. Nevertheless, it is evident that incentive compatibility is a desirable property of a system, for two reasons. First, the selfishness assumption for end-hosts is a pessimistic assumption for networked systems, and designing systems under this assumption limits the possible damage. Second, a system in which users can gain by rule-breaking may be seen as inherently unfair, which limits the adoption of the system. 
% With this work, we have made a first step in addressing these concerns regarding path-aware network architectures, which hopefully sparks further research into load-adaptive path selection and its impact on these architectures.

\bibliographystyle{plain}
\bibliography{bibliography}
\clearpage

\begin{appendix}

\section{Example of Stability}
\label{sec:model:example-stability}

The oscillation-prone system from Section~\ref{sec:model:example} is stable if a sufficient number of end-hosts anticipate the greedy strategy $\sigma_\mathrm{g}$ with an \emph{antagonist} strategy $\sigma_\mathrm{a}$.
An end-host adopting the antagonist strategy always selects the path with the higher perceived cost, speculating that the seemingly cheaper path will soon be overloaded by greedy-strategy players: \begin{equation}u_\mathrm{a}(\pi,t\,|\,\tilde\pi) = \begin{cases}1 & \text{if } c_{\pi}(t-T) > c_{\tilde\pi}(t-T)\\ 0 & \text{otherwise}\end{cases}\end{equation}
Conversely, $u_\mathrm{a}(\tilde\pi,t\,|\,\tilde\pi) = 1-u_\mathrm{a}(\pi,t\,|\,\tilde\pi)$.

In an oscillation-prone system with strategy profile $v = \{\sigma_\mathrm{g} \mapsto q, \sigma_\mathrm{a} \mapsto 1-q\}$ and initial imbalance $A_0 > 1/2$, the initial dynamics of the system are \begin{equation}
f_{\alpha}(t)=(A_0+q-1)e^{-rt} + (1-q).
\end{equation} 

For $q \leq 1/2$, we see that $f_{\alpha}(t) > f_{\beta}(t)$ for all $t \geq 0$, since $\lim_{t\rightarrow\infty} f_{\alpha}(t) = 1-q \geq 1/2$, $f_{\alpha}(0) = A_0 > 1/2$, and $f_{\alpha}(t)$ is monotonic. Using the definitions from \cref{sec:model:system}, the oscillation-prone system is \emph{stable} with $\Delta^* = 1-2q$ for all $q < 1/2$ and is \emph{stable at equal load} for $q = 1/2$.

\section{Example of PSS Equilibrium Analysis}
\label{sec:model:equilibrium:test}

In this section, we illustrate the calculation of strategy costs of the form set out in \cref{sec:model:nash} by investigating whether the strategies described in \cref{sec:model:example-stability} form PSS equilibria. Proving that a strategy profile is not a PSS equilibrium amounts to finding a deviant strategy that reduces an end-host's cost. Indeed, there exist such deviant strategies for the strategy profile $v$ with $v(\sigma_\mathrm{g}) = q$ and $v(\sigma_\mathrm{a}) = 1-q$ for all $q \in [0,1]$.

For the case $q \leq 1/2$, there is no inversion of link costs and
a deviant agent can always assume that $f_{\pi}(t) > f_{\tilde\pi}(t)$ if the agent perceives $f_{\pi}(t-T) > f_{\tilde\pi}(t-T)$. The best strategy given such a strategy profile thus consists of switching to the cheaper path~$\tilde\pi$ in a deterministic and immediate fashion, as in the greedy strategy $\sigma_\mathrm{g}$ presented in \cref{sec:model:example}. Every delay of switching simply translates into more time needlessly spent on a strictly more expensive path. As the greedy strategy $\sigma_\mathrm{g}$ allows an end-host to reduce its cost, $v(\sigma_\mathrm{g})$ would quickly rise from $q$ as more end-hosts adopt this strategy. Therefore, any strategy profile with $q \leq 1/2$ is not a PSS equilibrium.

For $q > 1/2$, the periodic dynamics are structured as
\begin{equation}f_{\alpha}(t) = 
            \begin{cases}
                (A+q-1)\cdot e^{-rt'} + 1-q & \text{if } \frac{t^+(t)}{W} \text{ is even},\\
                -(A+q-1)\cdot e^{-rt'} + q & \text{otherwise},
            \end{cases}
\end{equation} 
where $t'=t-t^+(t)$, 
\begin{equation}
W = \frac{\ln(2e^{rT}-1)}{r}, \quad \text{and} \quad A=\left(\frac{1}{2}-q\right)e^{-rT}+q.
\end{equation}
            
For showing that the antagonist strategy $\sigma_{\mathrm{a}}$ allows an end-host to improve its cost if $q \in (1/2, 1]$, we construct a mixed strategy $\sigma_{\mathrm{p}}(q')$. This strategy $\sigma_{\mathrm{p}}(q')$ plays the greedy strategy $\sigma_{\mathrm{g}}$ with probability $q'$ and the antagonist strategy $\sigma_{\mathrm{a}}$ with probability $1-q'$. We show that an end-host minimizes its cost by choosing $q' = 0$ given $q \in (1/2,1\big]$, i.e, the antagonist strategy $\sigma_{\mathrm{p}}(0) = \sigma_{\mathrm{a}}$ is the better strategy than the greedy strategy $\sigma_{\mathrm{p}}(1) = \sigma_{\mathrm{g}}$.

As mentioned in \cref{sec:model:nash}, the cost of a strategy in periodic oscillating systems is computed over a single periodic interval. For the dynamics above, it is even sufficient to calculate the strategy cost between two turning points $t^+_0$ and $t^+_1$, as the costs of the paths $\alpha$ and $\beta$ would simply be reversed in the subsequent turning-point interval. Without loss of generality, we thus operate on a turning-point interval $[t_0^+, t_1^+]$ during which path $\alpha$ is perceived to be the cheaper path and $f_{\alpha}(t_0^+) < f_{\beta}(t_0^+)$.

The time-dependent strategy cost $C(\sigma_p(q'),t)$ for the deviant agent is calculated based on a linear combination of the two path costs, weighted by $q'$: 
\begin{equation}
C(\sigma_p(q'),t) = \frac{1}{R}\int_t^{t+R} \left[q'\cdot c_{\alpha}(s) + (1-q')\cdot c_{\beta}(s)\right] \d{s}
\label{eq:eq-test:integral}
\end{equation}
We further assume $R \leq W$, as any choice of higher $R$ forces an agent to select a path that is sub-optimal during at least time $R-W$. Using this limitation, it is possible to derive a formula for the strategy cost $C(\sigma_p(q')|O)$ that is a linear function of $q'$,
\begin{equation}
C(\sigma_p(q')|O) = m\cdot q' + \gamma
\end{equation}
where $\gamma$ is constant w.r.t. $q'$ and the slope $m$ is 
\begin{equation}
\frac{R\big[(2q-1)(W-R)+\frac{2a}{r}(e^{-rW}+1)\big]+\frac{4a}{r^2}(e^{-rR}-1)}{RW}
\end{equation}
using the abbreviation $a = A+q-1$. The cost function steepness
is assumed to be~$p=1$, as the integral 
in~\cref{eq:eq-test:integral} is not tractable otherwise.

The slope $m$ can be shown to be positive for all $R  > 0$, $r \in [0,1]$, and $T \geq T(R)$, where $T(R)$ is such that $W = R$. Showing this property is feasible in a two-step proof, where we first show $m(T) > 0$ for $T = T(R)$ and $\partial/\partial T\ m(T) > 0$ for all $T > T(R)$.
The positiveness of $m$ implies that the minimum of the strategy cost $C(\sigma_p(q')|O)$ is achieved for $q' = 0$, i.e., the antagonist strategy $\sigma_\mathrm{a}$.

Given a strategy profile with $q > 1/2$, the adoption rate $q$ of the greedy strategy would thus quickly decrease in favor of the antagonist strategy $\sigma_\mathrm{a}$. Therefore, no strategy profile $v$ with $q > 1/2$ represents a PSS equilibrium.

\section{Proof of Observation \ref{obs:strategies:fischer}}
\label{sec:appendix:obs-fischer}

 We can numerically show that there exist oscillation-prone systems where the greedy strategy $\sigma_\mathrm{g}$ ensures a lower cost than an underdamped convergent strategy $\sigma_\mathrm{c}$. In fact, the oscillation-prone system $O$ assumed in \cref{fig:strategies:fischer-strategies} is such an oscillation-prone system where the strategy $\sigma_\mathrm{c}$ in an underdamped fashion does not yield the optimal cost. Using the definition of strategy cost introduced in \cref{sec:model:nash}, we calculate both $C(\sigma_\mathrm{c}|O)$ and $C(\sigma_\mathrm{g}|O)$.

In the calculation of $C(\sigma_\mathrm{c}|O)$, we choose $u(\pi,t\,|\,\pi_t)$ as defined in \cref{eq:strategies:switching-probability-fischer}. Furthermore, we can assume that $y(\pi_t|t) = f_{\pi_t}(t)$, because an agent applying strategy $\sigma_\mathrm{c}$ allocates its traffic in accordance with all other agents and its probability distribution of being on a certain path is equivalent to the general traffic distribution over the paths. As for the calculation of $C(\sigma_\mathrm{g}|O)$, we know that
\begin{equation}u(\pi,t\,|\,\tilde\pi) = \begin{cases}1 & \text{if } c_{\pi}(t-T) < c_{\tilde\pi}(t-T),\\0 &\text{otherwise,}\end{cases}\end{equation} 
and 
\begin{equation}y(\pi|t) = \begin{cases}1 & \text{if } c_{\pi}(t-T) = \min_{\pi'} c_{\pi'}(t-T),\\0 &\text{otherwise}.\end{cases}\end{equation}

\begin{figure}
    \centering
    \includegraphics[width=\linewidth]{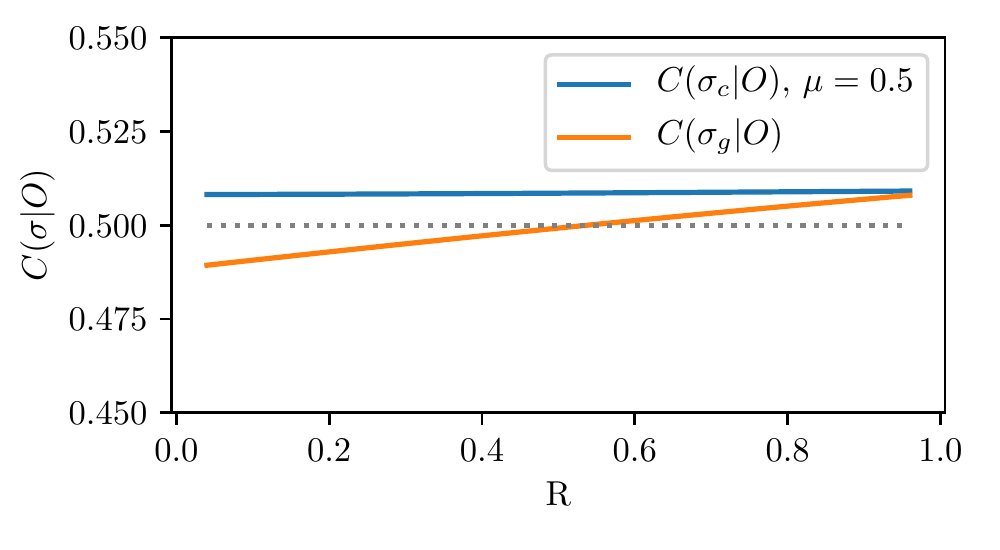}
    \vspace{-30pt}
    \caption{Example calculation illustrating that underdamped convergent strategy $\sigma_\mathrm{c}$ may be an inferior strategy (Environment: Oscillation-prone system $O = (\{\alpha,\beta\},r=1,p=1,T=2,A_0=1,v=\{\sigma_\mathrm{c}\mapsto1\})$).}
    \label{fig:strategies:underdamped-inferior}
\end{figure}

In \cref{fig:strategies:underdamped-inferior}, the comparison of strategy costs for $\sigma_\mathrm{c}$ and $\sigma_\mathrm{g}$ are shown for all $R \in [0,1]$ and the mentioned oscillation-prone system~$O$. Clearly, given the oscillation-prone system $O$ where agents universally apply an underdamped convergent strategy $\sigma_\mathrm{c}$, any single agent would have an incentive to switch to a greedy strategy $\sigma_\mathrm{c}$. The underdamped convergent strategy $\sigma_\mathrm{c}$ is thus not a PSS equilibrium.

\section{Proof of Observation \ref{obs:strategies:mate}}
\label{sec:appendix:obs-mate}

The flow-allocation vector $\mathbf{F}^{\sim}$ before projection is given by (using the abbreviation $f_{\pi}$ for $f_{\pi}(t-T)$) \begin{equation}\mathbf{F}^{\sim}=\begin{pmatrix}F_{\alpha}-\gamma\cdot c_{\alpha}\\F_{\beta}-\gamma\cdot c_{\beta}\end{pmatrix}.\end{equation}

The projection on the feasible allocation set is the intersection of the line describing the feasible set $F_{\beta}' = d - F_{\alpha}'$ and the line through $\mathbf{F}^{\sim}$ which is orthogonal to the feasibility line: 
\begin{equation}F_{\beta}' = F_{\alpha}' + \big(F_{\beta}-F_{\alpha}-\gamma(c_{\beta}-c_{\alpha})\big)\end{equation} 
This intersection is at $F_{\alpha}' = 1/2\cdot\big(d-F_{\beta}+F_{\alpha}+\gamma(c_{\beta}-c_{\alpha})\big)$. The change in an end-host's flow on path $\alpha$ is thus 
\begin{equation}F_{\alpha}'-F_{\alpha} = \gamma/2\cdot\big(c_{\beta}(t-T)-c_{\alpha}(t-T)\big).\end{equation} 
If path $\alpha$ appears to be the more expensive path, this change is performed by the re-evaluating end-hosts on path $\alpha$, and otherwise by the re-evaluating end-hosts on path $\beta$. Multiplying by the number of re-evaluating end-hosts thus yields the aggregate dynamics 
\begin{equation}\frac{\partial f_{\alpha}}{\partial t} = \begin{cases} r\cdot \frac{\gamma}{2} \cdot \Delta(t-T) \cdot f_{\alpha}(t) & \text{if } \Delta(t-T) \leq 0 \\ r\cdot \frac{\gamma}{2} \cdot \Delta(t-T) \cdot f_{\beta}(t) & \text{otherwise} \end{cases} \end{equation} where $\Delta(t-T)= c_{\beta}(t-T)-c_{\alpha}(t-T)$.

\section{CROSS Stability Analysis}
\label{sec:cross:stability}

To prove \cref{lem:cross:stability}, we show that stability at
approximately equal load arises given universal adoption
of path-selection strategy $\sigma_{\mathrm{C}}$, 
i.e., end-hosts use a path if they have a registration for that path
and only use a backup path in case of a path failure.

For stability at approximately equal load with parameter~$\epsilon$,
we assume that an end-host does not reallocate traffic at time $t$
if the imbalance between paths $\Delta(t-T) = |f_{\alpha}(t-T) - 
f_{\beta}(t-T)|$ is  less than $\epsilon$
and thus the perceived cost difference is too small to justify
path migration. If the imbalance~$\Delta(t)$ 
can be kept below $\epsilon$ for a period of length $T$,
i.e., $\Delta(t) < \epsilon$ for all $t \in [\tilde t, \tilde t+T)$,
there will be no reallocation
during the following interval $[\tilde t+T, \tilde t + 2T)$ and, by
extension, also none in all subsequent intervals. 

In any balancing trial with start $t_i$, there will result a 
traffic imbalance $\Delta(t_i) = |f_{\alpha}(t_i) - 
f_{\beta}(t_i)|$. This imbalance remains constant during time
$[t_i, t_i+T)$, as the end-hosts only perceive the imbalance
at time $t_i + T$. Thus, if $\Delta(t_i) < \epsilon$, stability
at approximately equal load is reached and enforcement
of the mechanism can be suspended. However, if 
$\Delta(t_i) \geq \epsilon$, stability is not achieved and the 
balancing trials are repeated until $\Delta(t_i) < \epsilon$.

Since an end-host selects each path with probability 1/2, the 
distribution of $f_{\alpha}(t_i)$ on
$[0,1]$ can be approximated with a normal distribution 
$\mathcal{N}$ possessing mean $\mu = 1/2$
and variance $\sigma^2$ that depends on the number of 
end-hosts. If $\Phi(f_{\alpha})$ is the CDF of $\mathcal{N}$, then 
the  probability that $\Delta(t_i) < \epsilon$ is 
$p_{<\epsilon} = \Phi((1+\epsilon)/2)-\Phi((1-\epsilon)/2) > 0$.
With an increasing number of balancing trials over time~$t$,
the probability that $\Delta(t_i) < \epsilon$ goes to 1
for $t\rightarrow\infty$. Therefore, for $t\rightarrow~\infty$,
it also holds that $\Delta(t) < \epsilon$, which is stability
at approximately equal load. \cref{lem:cross:stability}
thus holds.

Indeed, the CROSS mechanism eventually achieves 
stability at approximately equal load even without relying
on the computational puzzles mentioned in \cref{sec:cross:overview}.
However, it is desirable that oscillation can already be avoided
during the execution of the mechanism. In particular, 
if a balancing trial fails and $\Delta(t_i) \geq \epsilon$,
no oscillation should take place until the start of the next balancing trial, i.e., during time $[t_i+T,t_{i+1})$.
If the imbalance $\Delta(t_i)$ becomes visible to end-hosts
at time $t_i + T$, the end-hosts on path $\tilde\pi$ with a backup registration for path $\pi$ could migrate. However, since 
CROSS ensures that an end-host with a backup registration only uses 
its backup path in case of a path failure (see next section),
no migration takes place at all during $[t_i+T, t_{i+1})$.
Therefore, in absence of a path failure, the load distribution
remains constant during the whole duration $[t_i, t_{i+1})$
of a balancing trial.

\end{appendix}

\end{document}